\documentclass[usenatbib]{mnras}
\usepackage{graphicx}
\usepackage{amssymb}
\usepackage{amsmath}
\usepackage{bm}
\usepackage{hyperref}

\title[Radiative simulations of steady-state plasmoid reconnection]{Radiative kinetic simulations of steady-state relativistic plasmoid magnetic reconnection}

\author[Ortu\~no-Mac\'ias \& Nalewajko]{
Jos\'e Ortu\~no-Mac\'ias \& Krzysztof Nalewajko
\\
Nicolaus Copernicus Astronomical Center, Polish Academy of Sciences, Bartycka 18, 00-716 Warsaw, Poland
\\
\tt{jortuno@camk.edu.pl}
}

\begin{document}

\maketitle

\begin{abstract}
We present the results of 2D particle-in-cell (PIC) simulations of relativistic magnetic reconnection (RMR) in electron-positron plasma,
including the dynamical influence of the synchrotron radiation process, and integrating the observable emission signatures.
The simulations are initiated with a single Harris current layer with a central gap that triggers the RMR process.
We achieve a steady-state reconnection with unrestricted outflows by means of open boundary conditions.
The radiative cooling efficiency is regulated by the choice of initial plasma temperature $\Theta$.
We explore different values of $\Theta$ and of the background magnetisation $\sigma_0$.
Throughout the simulations, plasmoids are generated in the central region of the layer, and they evolve at different rates, achieving a wide range of sizes.
The gaps between plasmoids are filled by smooth relativistic outflows called minijets, whose contribution to the observed radiation is very limited due to their low particle densities.
Small-sized plasmoids are rapidly accelerated, however, they have lower contributions to the observed emission, despite stronger relativistic beaming.
Large-sized plasmoids are slow, but produce most of the observed synchrotron emission, with major part of their radiation produced within the central cores, the density of which is enhanced by radiative cooling.
Synchrotron lightcurves show rapid bright flares that can be identified as originating from mergers between small/fast plasmoids and large/slow targets moving in the same direction. 
In the high-magnetisation case, the accelerated particles form a broken power-law energy distribution with a soft tail produced by particles accelerated in the minijets.
\end{abstract}


\section{Introduction}
\label{sec_intro}

Signatures of non-thermal particle energy distributions are commonly observed in the high-energy astrophysical phenomena such as gamma-ray bursts \citep[e.g.,][]{2004RvMP...76.1143P}, pulsar wind nebulae \citep[e.g,][]{2012ApJ...749...26B} and blazars \citep[e.g.,][]{2016ARA&A..54..725M}.
What these diverse sources have in common is very broad photon spectra indicating broad non-thermal energy distributions of the radiating particles.

Blazars are a subclass of active galactic nuclei (AGN), in which a relativistic jet emerging from the supermassive black hole (SMBH) is aligned with our line of sight \citep{1995PASP..107..803U}.
They are extremely luminous sources of radiation that spans the entire electromagnetic spectrum, from radio up to very high energy $\gamma$-rays ($\sim$ TeV).
The extreme broadness and power-law appearance of the blazar spectra are signatures of an efficient non-thermal particle acceleration (NTPA) process.
Two major spectral components are typically observed with regular characteristics \citep{1998MNRAS.299..433F,ghi17}, with the low-energy one (radio to optical/UV/X-rays) interpreted as synchrotron emission, and the high-energy one (mainly $\gamma$-rays) due to either leptonic (inverse Compton) or hadronic processes \citep[e.g.,][]{2009ApJ...704...38S,2013ApJ...768...54B}.

The emission of blazars is also characterised by strong variability on timescales ranging from decades \citep[e.g.,][]{2016A&A...593A..91A,2017ApJ...837..127G} to minutes \citep[e.g.,][]{2007ApJ...664L..71A,2007ApJ...669..862A,2019ApJ...877...39M}.
The shortest variability timescales are much shorter than the light crossing time ($\sim$ hours) of the gravitational radius of the SMBH contained in these type of AGN ($M_{BH} \sim 10^9 M_{\odot}$).
Such short variability time scales and high luminosities require very compact emitting regions within the jet \citep{2007ApJ...671L..29L,2008MNRAS.384L..19B,2008MNRAS.390..371K}.

Compact regions containing relativistic particles and magnetic fields have been often invoked to explain the multi-wavelength emission of blazars \citep[e.g.,][]{1998A&A...333..452K,1999MNRAS.306..551C}.
Theoretical models for the formation of collimated jets with relativistic velocities require relativistically magnetised environments\footnote{In general, magnetisation parameter is defined as $\sigma = B^2 / (4\pi w)$, where $B$ is the magnetic field strength and $w$ is the relativistic enthalpy density of the matter.} \citep{1977MNRAS.179..433B,1992ApJ...394..459L,2007MNRAS.380...51K}.
In such environments, shock waves are weak structures that do not provide enough energy dissipation to power the observed emission \citep{2015MNRAS.450..183S}.

Magnetic reconnection is a dissipation mechanism that is most efficient in strongly magnetised plasmas.
During the reconnection, free magnetic energy is transferred to the thermal heating and bulk acceleration of plasma, as well as to the NTPA process \citep{2009ARA&A..47..291Z}.
Numerous arguments point to the RMR as the process responsible for the multi-wavelength and multi-timescale variability emission of blazars \citep{2015MNRAS.450..183S}.
A basic requirement for reconnection to occur in relativistic jets is to have local inversions of the magnetic field lines, which may be caused by internal jet instabilities, in particular the current-driven kink modes \citep{1998ApJ...493..291B,2006A&A...450..887G,2018PhRvL.121x5101A,2019ApJ...884...39B}, or by global magnetic polarity reversals \citep{1997ApJ...484..628L,2019MNRAS.484.1378G}.

Traditional models of magnetic reconnection \citep{1958IAUS....6..123S,1957JGR....62..509P,1964NASSP..50..425P} have been adapted to the relativistic regime by \cite{2003ApJ...589..893L} and \cite{2005MNRAS.358..113L}. The latter work became a basis for the minijets model of relativistic bulk outflows driven by the RMR \citep{2009MNRAS.395L..29G,2011MNRAS.413..333N}.
Other works emphasised the fundamental role of plasmoids (or magnetic flux tubes) that arise spontaneously in sufficiently long and thin current layers due to the tearing instability \citep{2007PhPl...14j0703L,2013MNRAS.431..355G}.
Plasmoids trap energised particles and evacuate them together with the reconnected magnetic field from the active magnetic X-points, enhancing the overall reconnection rate\footnote{Inflow velocity of the background plasma.} \citep{2010PhRvL.105w5002U}.

The NTPA mechanism in the case of RMR has been studied extensively by means of kinetic particle-in-cell (PIC) numerical simulations.
In the most basic case of electron-positron pair plasma,
it has been established that NTPA produces power-law energy distributions of particles $N(E) \propto E^{-p}$ with localised power-law indices as hard as $p \simeq 1$ in the limit of $\sigma \gg 1$ \citep{2001ApJ...562L..63Z,2014ApJ...783L..21S,Guo14,2016ApJ...816L...8W}, converging however to $p \simeq 2$ in the long term \citep{2018MNRAS.481.5687P}.

Most PIC simulations of the RMR are initiated from Harris-type current layers contained in periodic domains.
In the relatively small domains, this leads to the cancellation of reconnection outflows momenta and to the formation of artificially large single plasmoids.
An alternative approach is to use open boundaries that allow outgoing particles to escape freely and absorb the outgoing Poynting flux \citep{2006PhPl...13g2101D}.
This approach allows to investigate magnetic reconnection as a sustained steady-state process with unimpeded outflows approaching the terminal Alfven velocity and continuous generation and evolution of plasmoids \citep{2007PhPl...14g2303D}.

The open-boundary approach has been first applied to the case of RMR by \cite{2016MNRAS.462...48S}, who investigated the statistical properties of plasmoids, determined their size distribution as a power-law, and demonstrated an anti-correlation between plasmoid growth and bulk acceleration.
The corresponding predictions for the variability of non-thermal radiation were applied to explain the observational characteristics of blazar flares \citep{2016MNRAS.462.3325P}.
Based on the plasmoid scaling laws, a stochastic model for the evolution of plasmoid chains was developed by \cite{2018MNRAS.475.3797P}.
The results of these PIC simulations were also post-processed to calculate light curves of synchrotron and inverse Compton emission as would be observed if the reconnection region was located in a relativistic jet \citep{2019MNRAS.482...65C,2020MNRAS.492..549C}.
A semi-analytical model of broad-band emission from reconnection plasmoids in the context of blazar flares was also developed by \cite{2019MNRAS.486.1548M}.

Other numerical studies of the RMR were performed in the context of $\gamma$-ray flares from the Crab Nebula, taking into account the effect of radiation reaction on individual particles \citep[e.g.,][]{2013ApJ...770..147C,2014ApJ...782..104C}.
Recent PIC simulations which included the effects of synchrotron and inverse Compton cooling have shown that the particle acceleration can be decreased in comparison with previous non-radiative results \citep{2018MNRAS.481.4342N,2019ApJ...870...49S,2019ApJ...877...53H,2019arXiv190808138S}.
We may therefore expect that evolution of plasmoids will be affected by radiative cooling.

In this work, we present the results of PIC simulations of the steady state RMR process allowed by the open boundaries under strong radiative energy losses due to the synchrotron radiation reaction.
We investigate in detail the evolution of individual plasmoids under different levels of radiative cooling, and calculate accurate lightcurves of synchrotron radiation.
We demonstrate a connection between rapid radiation flares and tail-on mergers of small/fast and large/slow plasmoids.

Plan of this work:
\S\ref{sec_setup} simulations setup;
\S\ref{sec_methods} analysis methods;
%\S\ref{sec_results} results;
\S\ref{sec_res_init} initial sequence;
\S\ref{sec_res_xymap} evolved reconnection layer;
\S\ref{sec_res_xtmap} spacetime diagrams of the current layer;
\S\ref{sec_res_dens_temp} particle density and mean energy distributions;
\S\ref{sec_res_plasm} analysis of individual plasmoids;
\S\ref{sec_res_part} acceleration and cooling of selected individual particles;
\S\ref{sec_res_lc} synchrotron light curves;
\S\ref{sec_res_enecons} global energy conservation;
\S\ref{sec_res_spectra} energy distributions of particles and photons;
\S\ref{sec_disc} discussion;
\S\ref{sec_conc} conclusions.

\section{Simulation setup}
\label{sec_setup}

We make use of a custom version of the Particle-in-Cell (PIC) code {\tt Zeltron} \citep{2013ApJ...770..147C}.
We perform 2D simulations of relativistically magnetised pair plasma in a fixed tall domain of dimensions $L_x \times L_y$ with $L_y = 4L_x$, within the coordinate ranges $0 < x < L_x$ and $-L_y/2 < y < L_y/2$.

Our simulations are initiated from an equilibrium configuration $\bm{B}(t=0) = \bm{B}_{\rm ini}$ and $\bm{E}(t=0) = \bm{E}_{\rm ini} = 0$ that involves a single Harris-type current layer 
placed in the middle of the computational domain:
\begin{equation}
\label{eq:B_x}
B_{\rm ini,x} = -B_0\tanh{(y/\delta)}\,,
\end{equation}
where \( B_0 \) is the characteristic value of magnetic field strength, and \( \delta \) is the Harris layer half-thickness.
We include no initial guide field, hence $B_{\rm ini,z} = 0$.
The magnetic field gradient across the Harris layer is supported by the electric current and pressure provided by the drifting population of particles that is characterised by the Maxwell-J{\"u}ttner energy distribution with dimensionless temperature \( \Theta_{\rm d} = k_{\rm B}T_{\rm d}/(m_{\rm e}c^2) \), Lorentz-boosted with the drift velocity \( \beta_{\rm d} = v_{\rm d}/c = 0.3 \), and number density
\begin{equation}
\label{eq:n_d}
n_{\rm d}(y) = n_{\rm d,0}\cosh^{-2}(y/\delta)\,,
\end{equation}
where $n_{\rm d,0} = \gamma_{\rm d} B_0^2 / (8\pi\Theta_{\rm d}m_{\rm e}c^2)$, and $\gamma_{\rm d} = (1-\beta_{\rm d}^2)^{-1/2}$ is the drift Lorentz factor.
The Harris layer thickness is related to the nominal plasma gyroradius $\rho_0 = \Theta_{\rm d}m_{\rm e} c^2/(e B_0)$ as $ \delta = 2\rho_0/(\gamma_{\rm d}\beta_{\rm d})$ \citep{2003ApJ...591..366K}.

The current layer is immersed in a background plasma with the Maxwell-J{\"u}ttner energy distribution with dimensionless temperature $\Theta_{\rm b} = k_{\rm B}T_{\rm b}/(m_{\rm e}c^2) = \Theta_{\rm d} \equiv \Theta$, and number density $n_{\rm b}$ that is determined by the assumed value of the magnetisation $\sigma_0 = B_0^2/(4\pi n_{\rm b} \Theta_{\rm b} m_{\rm e}c^2)$.
The corresponding initial density contrast can be expressed as
$n_{\rm d,0} / n_{\rm b} = \gamma_{\rm d}\sigma_0/2$.

Our choice of drift velocity means that the current layer is relatively thick, and hence stable to the tearing modes. We experimented with higher values $\beta_{\rm d} \simeq 0.5$, which results in spontaneous formation of slowly evolving plasmoids. In order to speed up the evolution of the current layer, we follow \cite{2016MNRAS.462...48S} in triggering fast magnetic reconnection by placing a narrow gap ($\Delta x = 8\delta$) in the middle of the current layer. Within the trigger gap, the distribution of drifting particles is initialised with temperature reduced by factor 10, and with no drift velocity.

The dimensions of our computational domain are up to \(N_x \times N_y = 4608 \times 18432\) cells with equal spatial resolution for both dimensions \( {\rm d}x = {\rm d}y = \rho_0/3\),
which results in the physical scales of \( L_x \times L_y = (1536 \times 6144)\rho_0\).
The temporal resolution is \({\rm d}t = 0.9\,{\rm d}t_{\rm CFL}\), where ${\rm d}t_{\rm CFL} = [({\rm d}x)^{-2} + ({\rm d}y)^{-2}]^{-1/2}/c$ is the Courant-Friedrichs-Lewy (CFL) time step.
The initial distributions of background and drifting electrons/positrons are represented on average by 64 macroparticles per cell per species.

Our computational domain is open at left/right boundaries, where outgoing particles are removed, and fresh particles are injected at every timestep at a fixed rate
%$\dot{N}_{\rm inj} = \Delta N_{\rm inj}/[\Delta(ct)\,\Delta y]$
calculated to maintain the initial number density of both background and drifting particles.
At every timestep, we inject a fixed number of particles along each boundary ${\rm d}N_{\rm inj}/{\rm d}(ct) = (\left<\beta\right>/4){\rm d}N_{\rm ini}/{\rm d}x$, where ${\rm d}N_{\rm ini}$ is the number of initial particles in a single grid column, and $\left<\beta\right> = \left<|\bm\beta|\right> = \left<v\right>/c \simeq 1$ is the mean particle dimensionless velocity
module\footnote{The factor $\left<\beta\right>/4$ is derived from the rate at which particles of uniform density and isotropic momentum distribution $N(\bm{u}) = N f(u) g(\mu)$ cross the boundary.
Here, $f(u)$ is the Maxwell-J{\"u}ttner distribution (normalised to unity) of momentum module $u = |\bm{u}| = \gamma\beta$ with $\gamma = (1-\beta^2)^{-1/2} = (1+u^2)^{1/2}$ the Lorentz factor,
and $g(\mu)$ is the uniform distribution of the cosine parameter $\mu = u_x/u = \beta_x/\beta$.
For isotropic target particle distribution, $g(\mu) = 1/2$ for $\mu\in[-1:1]$, hence $\int_{-1}^{1}{\rm d}\mu\,g(\mu) = 1$.
For injected particles, this distribution is truncated to particles inflowing to the domain -- $\mu\in(0:1]$ in case of left boundary, and $\mu\in[-1:0)$ in case of right boundary.
For particles of momentum $\bm{u}$ we find ${\rm d}N(\bm{u})/{\rm d}(ct) = \beta_x\,{\rm d}N(\bm{u})/{\rm d}x$, integrating this over $\bm{u}$ we find ${\rm d}N/{\rm d}(ct) = {\rm d}N/{\rm d}x\,\int_{0}^{\infty}{\rm d}u\,\beta f(u)\,\int_{0}^{1}{\rm d}\mu\,\mu g(\mu) = {\rm d}N/{\rm d}x\,\left<\beta\right>\,(1/4)$.}.
The injected particles are located at fixed distance just within each boundary ($x_0 = \Delta x/100$ in case of left boundary; $x_0 = L_x - \Delta x/100$ in case of right boundary).
Their distribution along the $y$ coordinate is proportional to the distribution of initial particles.
The injected particles are likewise divided into background and drifting populations.
A Lorentz boost is applied to the drifting particles in order to match the current density $j_{z,\rm ini}$ of the initial Harris layer (otherwise, the initial current layer would evolve strongly from the boundaries inwards).

Within the left/right boundaries we place the field-absorbing layers of thickness \( \Delta_{\rm abs} = 30\,{\rm d}x \).
In the absorbing layers, in addition to the standard time advance of magnetic and electric fields, at every time step we perform the following operation:
\begin{eqnarray}
\bm{B}(x) &\to& \bm{B}(x) + \lambda(x)\left[\bm{B}_{\rm ini}(x) - \bm{B}(x)\right]\,,
\label{eq:B_abs}
\\
\bm{E}(x) &\to& \bm{E}(x) + \lambda(x)\left[\bm{E}_{\rm ini}(x) - \bm{E}(x)\right]\,,
\label{eq:E_abs}
\end{eqnarray}
where $\lambda(x) = 0.5(|x - x_{\rm abs}|/\Delta_{\rm abs})^3$,
and \( x_{\rm abs} \) is the position of the absorbing layer inner edge.
For the top/bottom boundaries, we apply periodic conditions for the particles and fields, while only the $z$-component of the electric field is absorbed using the above rule.

We apply the synchrotron radiation reaction to every particle at every timestep\footnote{We apply a restriction such that particles cannot lose more than half of their energy or reverse their momentum over a single time step.}, following the prescription of \cite{2013ApJ...770..147C}: 
\begin{equation}
\label{eq:Psynchrotron}
\frac{\partial\bm{u}}{\partial t} = -\frac{P_{\rm syn} \bm{u} }{\gamma m_{\rm e}c^2}, 
  \; \;\;\;
P_{\rm syn} = \frac{\sigma_{\rm T}c}{4\pi}\left[(\gamma \bm{E} + \bm{u}\times\bm{B})^2 -   (\bm{E}\cdot\bm{u})^2\right]\,,
\end{equation}
where $\sigma_{\rm T} = (8\pi/3)r_{\rm e}^2$ is the Thomson cross section with $r_{\rm e} = e^2/m_{\rm e}c^2$ the classical electron radius, and $\bm{u} = \gamma \bm\beta$ is the dimensionless momentum of a particle with dimensionless velocity $\bm\beta = \bm{v}/c$ and Lorentz factor $\gamma = (1-\beta^2)^{-1/2} = (1 + u^2)^{1/2}$.
Noting that for $\gamma \gg 1$ we have $u \simeq \gamma$, the radiative cooling rate can be estimated as:
\begin{equation}
\frac{\partial \gamma}{\partial t} =
\frac{\bm{u}}{\gamma}\cdot\frac{\partial\bm{u}}{\partial t}
\simeq
-\frac{P_{\rm syn}}{m_{\rm e} c^2}\,.
\end{equation}
In the limit of zero electric field and isotropic particle distribution, Eq. (\ref{eq:Psynchrotron}) reduces to \( P_{\rm syn} = (4/3)\sigma_{\rm T} cu^2U_{B,0} \), where \( U_{B,0} = B_0^2/8\pi\) is the initial background magnetic energy density.
Taking into account that for the initial Maxwell-J{\"u}ttner distribution we have \(\left<\gamma\right> \simeq 3\Theta\) and \( \left<\gamma^2\right> \simeq 12\Theta^2 \), we define the nominal cooling length as \citep{NalYuaChr18}: 
\begin{equation}
\label{eq_lcool}
l_{\rm cool} = c\tau_{\rm cool} = \frac{\langle\gamma\rangle}{\langle \vert {\rm d}\gamma /c{\rm d}t\vert \rangle} \simeq \frac{\langle\gamma\rangle}{\langle\gamma^2\rangle} \frac{3m_{\rm e}c^2}{4\sigma_{\rm T}U_{\rm B,0}} \simeq \frac{(3\pi/2)e}{\sigma_{\rm T} \Theta^2 B_0} \rho_0\,.
\end{equation}
This can be compared with the radiation reaction limit on electron Lorentz factor $\gamma_{\rm rad} = (3\rho_0/2\Theta r_{\rm e})^{1/2}$ that can be achieved locally under crossed electric and magnetic fields of equal strengths $E = B$ \citep[e.g.,][]{2014ApJ...782..104C}.
We find a relation between $l_{\rm cool}$ and $\gamma_{\rm rad}$ by eliminating $B_0$:
\begin{eqnarray}
\frac{\left<\gamma\right>}{\gamma_{\rm rad}} &=& \left(\frac{27}{8}\frac{\rho_0}{l_{\rm cool}}\right)^{1/2}\,.
\label{eq_gamma_rad}
\end{eqnarray}
The total synchrotron power emitted in all directions per volume element of the initial background plasma is given by:
\begin{equation}
\label{eq_Psyn0}
\mathcal{P}_{\rm syn,b,0} \equiv \frac{{\rm d}E_{\rm syn}}{{\rm d}x\,{\rm d}y\,{\rm d}t} = 16c\sigma_{\rm T}n_{\rm b}\Theta^2U_{\rm B,0}\,.
\end{equation}

We also collect the synchrotron radiation spectra that would be measured by observers placed at the left and right side of the simulation domain.
The total spectrum is calculated as the sum of contributions from all macroparticles present in the domain \citep[see][and references therein]{NalYuaChr18}.
\begin{equation}
L_{\rm syn}(\nu) = \frac{\sqrt{3}e^2}{c} \sum_{e^+ e^-} N_{\rm e,1}F(\xi)\Omega_1\,,
\label{eq_Lsyn}
\end{equation}
where \(F(\xi) = \xi \int_{\xi}^{\infty}K_{5/3}(x){\rm d}x \),
\(\xi = 4\pi\nu/3\gamma^2\Omega_1\),
\(\Omega_1 = (e/m_{\rm e}c)\left|(\bm{E} + \bm{n} \times \bm{B}) \times \bm{n}\right|\),
\( \bm{n}=\bm{v}/|\bm{v}|\), \(N_{\rm e,1}\) is the number of charged particles (either electrons or positrons) represented by a single macroparticle, and \(K_{5/3}\) is the modified Bessel function of the second kind. 
A characteristic synchrotron frequency for monoenergetic electrons of Lorentz factor $\gamma$ can be derived from the isotropic synchrotron kernel\footnote{ Defined as $\mathcal{R}(\tilde\xi) = \tilde\xi^2\{K_{1/3}(\tilde\xi)K_{4/3}(\tilde\xi) - (3/5)\tilde\xi[K_{4/3}^2(\tilde\xi) - K_{1/3}^2(\tilde\xi)]\}$, where $\tilde\xi = 2\pi\nu/3\gamma^2\Theta\omega_0$ \citep{1986A&A...164L..16C}. Note that $\tilde\xi\,\mathcal{R}(\tilde\xi) \propto \nu\,F(\nu)$ peaks at $\tilde\xi_0 \simeq 0.575$.}:
\begin{equation}
\nu_{\rm syn}(\gamma) \simeq (3/2\pi)\tilde\xi_0\gamma^2\Theta\omega_0 \equiv \left(\frac{\gamma}{\Theta}\right)^2\nu_{\rm syn0}\,.
\label{eq_nu_syn0}
\end{equation}
%
%\begin{equation}
%\nu_{\rm syn}(\gamma) = (3/4\pi)\Theta\omega_0\gamma^2\,.
%\end{equation}
%
%Using the relation $\left<\gamma^2\right> = 12\Theta^2$, we obtain the nominal synchrotron frequency of the initial background plasma:
%
%\begin{equation}
%\nu_{\rm syn,b} = (9/\pi)\Theta^3\omega_0\,.
%\end{equation}
%
%If we instead substitute $\gamma = \gamma_{\rm rad}$, we obtain the magnetohydrodynamical (MHD) synchrotron frequency limit \citep[e.g.,][]{1983MNRAS.205..593G,2011ApJ...737L..40U}:
%
%\begin{equation}
%\nu_{\rm syn,max} = (9/8\pi)(c/r_{\rm e})\,.
%\label{eq_nu_syn_max}
%\end{equation}

We performed several large simulations for different values of background magnetisation $\sigma_0$ and initial particle distribution temperature $\Theta$.
In Table \ref{tab:param}, we report these parameters, as well as the corresponding ratios $l_{\rm cool}/L_x$ and $\left<\gamma\right>/\gamma_{\rm rad}$.
In our strong-cooling case {\tt s10Th}, $l_{\rm cool}$ is comparable to $L_x$, while the initial mean particle energy $\left<\gamma\right>$ is factor $\simeq 25$ below $\gamma_{\rm rad}$.

\begin{table}
\caption{Designations and key parameter values for our largest simulations performed in the domain of physical width \(L_x = 1536\rho_0\). Common parameters include: the initial background magnetic field strength $B_0 = 1\;{\rm G}$, and the initial dimensionless drift velocity $\beta_{\rm d} = 0.3$. The nominal cooling length $l_{\rm cool}$ is calculated from Eq. (\ref{eq_lcool}), and the radiation reaction limit ratio $\left<\gamma\right>/\gamma_{\rm rad}$ is calculated from Eq. (\ref{eq_gamma_rad}).}
\centering
\begin{tabular}{l r r r r r}
\hline\hline
name & \(\sigma_0\) & $\Theta$ & $l_{\rm cool}/L_x$ & $\left<\gamma\right> / \gamma_{\rm rad}\) \\
\hline
{\tt s10Tl} & 10 &    \(2\cdot10^5\) &  55 & 0.006 \\
{\tt s10Tm} & 10 &    \(5\cdot10^5\) & 8.9 & 0.016 \\
{\tt s10Th} & 10 & \(1.25\cdot10^6\) & 1.4 & 0.039 \\
{\tt s50Tm} & 50 &    \(5\cdot10^5\) & 8.9 & 0.016 \\
\hline\hline
\end{tabular}
\label{tab:param}
\end{table}

\section{Analysis Methods}
\label{sec_methods}

\subsection{Plasmoids identification}
\label{sec_meth_plasm}

\begin{figure*}
\includegraphics[width=\textwidth]{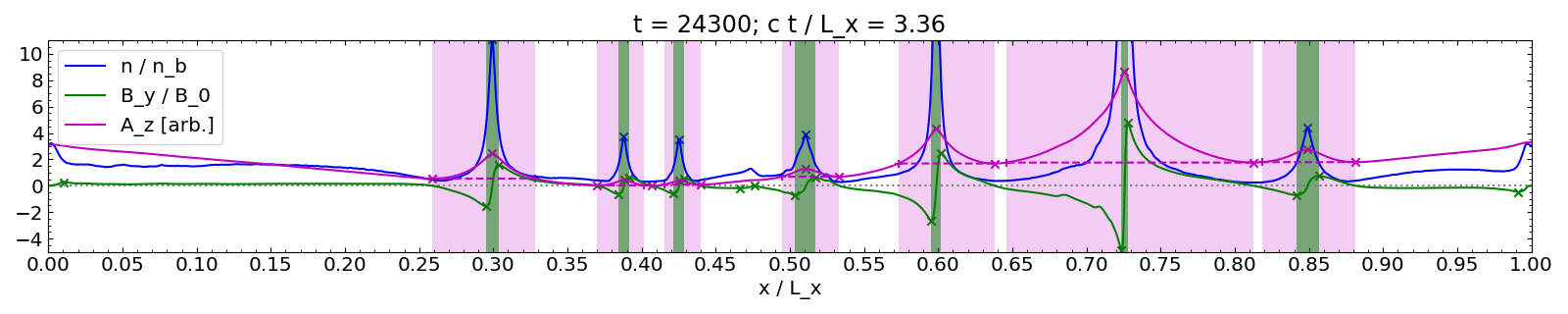}
\caption{Profiles of particle density (blue line), magnetic field component $B_y$ (green line), and magnetic vector potential component $A_z$ (magenta line) measured along the reconnection layer for a selected moment of simulation {\tt s10Tm} (cf. Figure \ref{fig:xymaps}).
The green vertical stripes mark the horizontal limits of plasmoid cores, and the light magenta areas mark the plasmoid layers.}
\label{fig:xprof}
\end{figure*}

We describe here an automated procedure that we use for identifying individual plasmoids along the reconnection layer.
During the simulations, we frequently save one-dimensional $x$-profiles of various parameters along the reconnection layer; more precisely, we extract these data from narrow rectangular stripes extending over $0 < x < L_x$ and $|y| < \delta/2$.
For each grid cell along the $x$-axis, we average the data over all grid cells along the $y$-axis that fit within the stripe.
The $x$-profiles were then Gaussian-smoothed with the standard deviation of $3\,{\rm d}x$. An example of such profiles is shown in Figure \ref{fig:xprof}.

We applied the {\tt scipy.signal.find\_peaks} algorithm to identify the maxima of particle number density $n$ (with the minimum prominence parameter of $1.5n_{\rm b}$), the minima and maxima of magnetic field component $B_y$ (with the minimum prominence of $0.02B_0$), and the minima and maxima of the magnetic vector potential component $A_z$ (with the minimum prominence of $0.22 B_0\rho_0$).
Next, we searched for all ordered pairs consisting of a $B_y$ minimum followed (to the right) by a $B_y$ maximum. For each such pair, we checked whether between those two points there is exactly one maximum of $n$ and one maximum of $A_z$. We also require that the distance between the $n$ maximum and the $A_z$ maximum is no more than $4\,{\rm d}x$.
Only the structures that satisfied all of the above conditions were classified as proper plasmoids.
We distinguish two structures within each plasmoid, a \emph{core} delimited by the minimum and maximum peaks of \( B_y \) (green stripes in Figure \ref{fig:xprof}), and a \emph{layer} where $A_z$ exceeds a threshold level equal to the higher of the closest $A_z$ minima located on both sides of the core and excluding the core itself (light-magenta areas in Figure \ref{fig:xprof}; this is consistent with the definition of plasmoid limits adopted by \citealt{2016MNRAS.462...48S}).
Qualitatively, the plasmoid cores are compact structures of high particle density, while the plasmoid layers are extended structures of relatively low particle density and closed magnetic field lines.

\subsection{Local reference frames}
\label{sec_meth_frame}

We convert certain parameters to their values in the local reference frames. As discussed in \cite{2018MNRAS.473.4840W}, in relativistically hot fluid there are two alternative ways to define a `co-moving' reference frame: (1) zero-current (Eckart) frame $\mathcal{O}'$ of velocity $\bm\beta_1 = \left<\bm\beta\right>$; and (2) zero-momentum (Landau) frame $\mathcal{O}''$ of velocity $\bm\beta_2 = \left<\gamma\bm\beta\right> / \left<\gamma\right>$. These averages are calculated locally for all particles (electrons and positrons) in a given grid cell as $\left<a\right> = (\sum_ia_in_i)/(\sum_in_i)$, where $n_i$ are the multiplicities of individual macroparticles such that $\sum_in_i \equiv n$ is the local particle number density.
For example, we calculate the electric field component in the Eckart frame as:
\begin{equation}
E_z' = \Gamma_1 \left(E_z + \left<\beta_x\right> B_y - \left<\beta_y\right> B_x\right)
\end{equation}
where $\Gamma_1 = (1 - \left<\beta_x\right>^2 - \left<\beta_y\right>^2)^{-1/2}$ is the Eckart bulk Lorentz factor.
On the other hand, we calculate the mean particle energy in the Landau frame as $\left<\gamma''\right> = \left<\gamma\right> / \Gamma_2$,
where $\Gamma_2 = (1 - \beta_{2,x}^2 - \beta_{2,y}^2)^{-1/2}$ is the Landau bulk Lorentz factor.

%\clearpage
\section{Results}
\label{sec_results}

\subsection{Initial sequence}
\label{sec_res_init}

\begin{figure*}
\includegraphics[width=\textwidth]{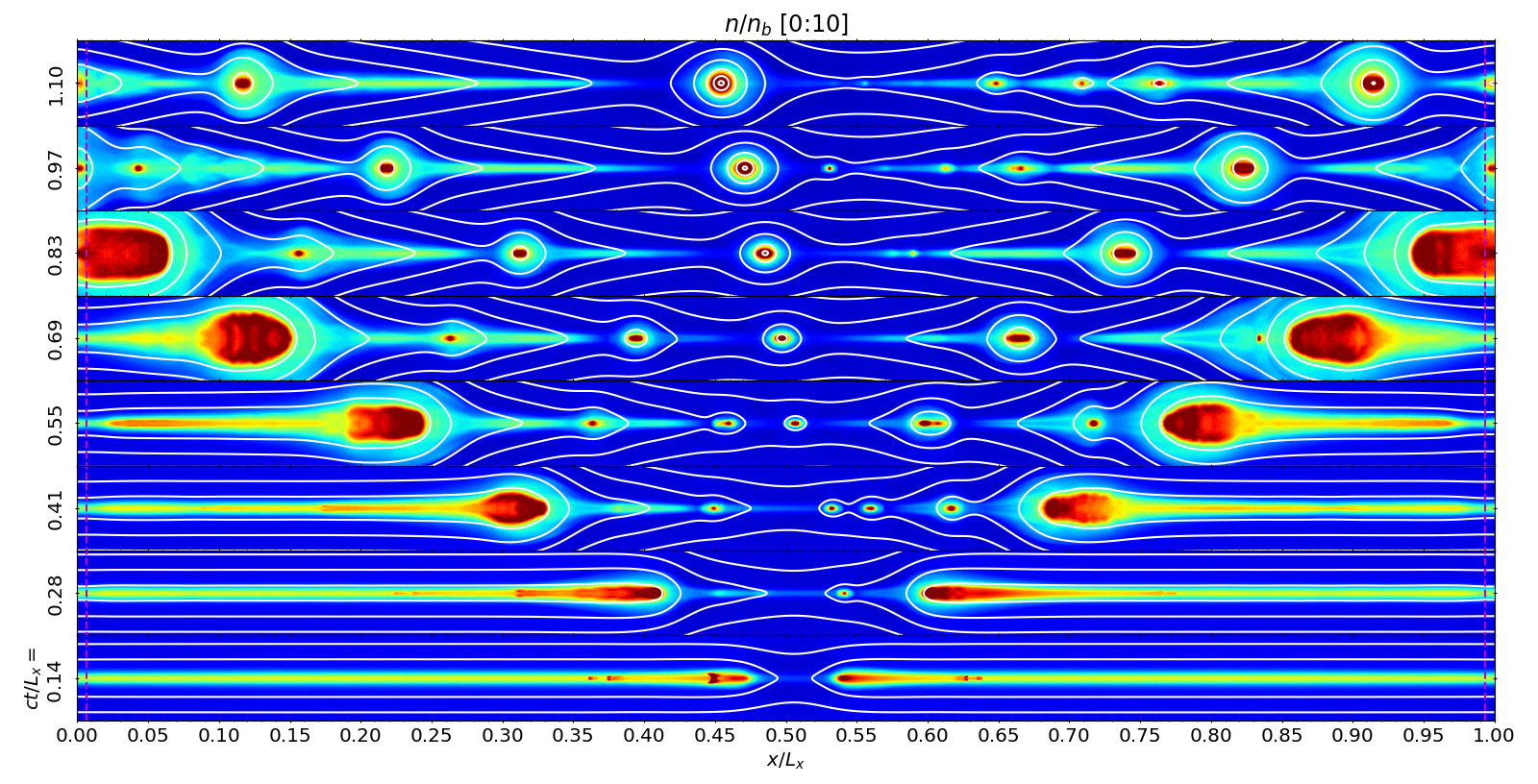}
\caption{Initial sequence of simulation {\tt s10Tm}, presenting the logarithm of particle number density $n / n_{\rm b}$ (see the top panel of Figure \ref{fig:xymaps} for the colour scale) and the magnetic field lines (solid white lines). The dashed magenta lines indicate the limits of the field-absorbing boundary layers.}
\label{fig:xymap_init}
\end{figure*}

Figure \ref{fig:xymap_init} shows the initial evolution of the reconnection layer in our reference simulation {\tt s10Tm}.
Each panel shows a particle number density map and magnetic field lines for a central region of our computational domain (only a $1/64$ fraction of its total vertical extent), in which the current layer is contained.
A centrally located gap in the current layer expands rapidly sideways towards the left/right boundaries, the initial drifting plasma component is pulled towards the boundaries by the tension of closed magnetic field lines.
Magnetic reconnection is triggered in the central low-density region, sustained by a thin current layer that is unstable to tearing modes which lead to the emergence of plasmoids.

At $ct \simeq L_x$, the swept-up fronts of the initial drifting plasma leave the left/right boundaries. From that moment on, we can describe the simulated reconnection process as steady-state, with newer plasmoids continuously generated around the centre of the layer, and older plasmoids escaping through the left/right boundaries.
The upper panels of Figure \ref{fig:xymap_init} demonstrate how our implementation of open boundaries works when large dense structures move across them.

We also find that the initial trigger gap induces a quasi-circular wave that propagates in all directions.
The horizontal wave fronts are effectively absorbed by the left/right boundaries, while the vertical wave fronts reach the top/bottom boundaries by $c t \simeq 2 L_x$ and are partially absorbed there, some very weak reflections return to the reconnection layer by about $c t \simeq 4L_x$ without a significant effect.

\subsection{Evolved reconnection layer}
\label{sec_res_xymap}

\begin{figure*}
\includegraphics[width=\textwidth]{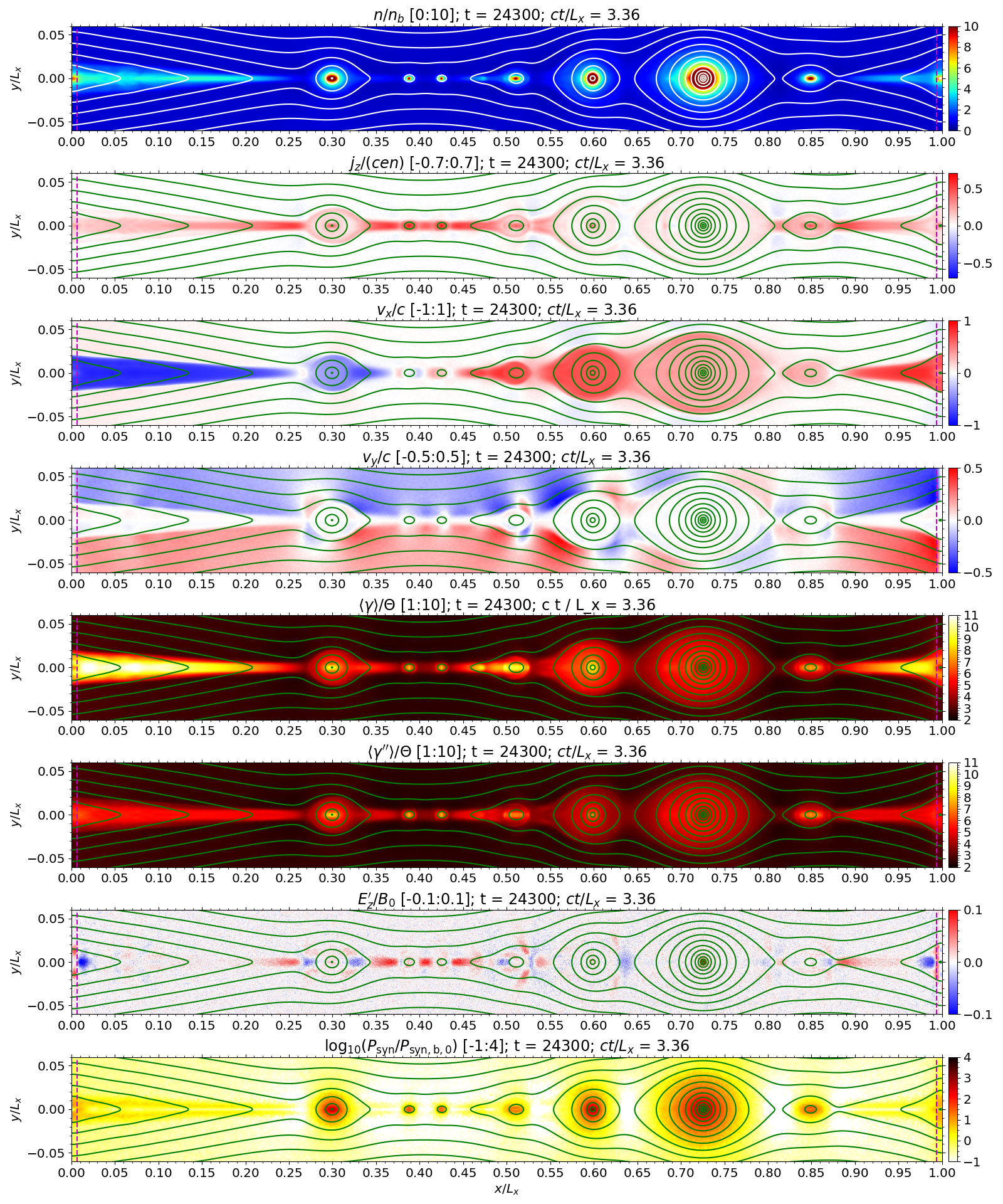}
\caption{Selected evolved state of simulation {\tt s10Tm} (the same as shown in Figure \ref{fig:xprof}). From the top, the panels show: (1) particle number density $n / n_{\rm b}$ in log scale, (2) out-of plane electric current per particle number density $j_z/(cen)$, (3) Landau velocity component along the current layer $v_x / c$, (4) Landau velocity component across the current layer $v_y / c$, (5) mean particle Lorentz factor measured in the simulation frame $\left<\gamma\right> / \Theta$, (6) mean particle Lorentz factor measured in the Landau co-moving frame $\left<\gamma''\right> / \Theta$, (7) out-of-plane electric field component measured in the Eckart co-moving frame $E_z'$, (8) total synchrotron power $E_{\rm syn}$ in log scale. The solid white lines show the magnetic field lines, and the dashed magenta lines indicate the limits of the field-absorbing boundary layers.}
\label{fig:xymaps}
\end{figure*}

Figure \ref{fig:xymaps} presents a selected evolved state of our reference simulation {\tt s10Tm} illustrated with multiple parameters. In this snapshot centred at the reconnection midplane, we observe diverse substructures.
The main current layer is observed at $0.34 < x / L_x < 0.49$, characterised by high values of the specific electric current $|j_z| / j_{\rm max}$ (electric current density $j_z$ close to its maximum value $j_{\rm max} = cen$) despite low particle number density $n$, low bulk velocity $v_x$, and strong electric field $E_z'$ (except for the two minor plasmoids located at $x \simeq 0.39 L_x$ and $x \simeq 0.425 L_x$).

To the left of the main current layer, behind a medium-sized plasmoid at $x \simeq 0.3 L_x$, we find a relativistically fast reconnection outflow (strongly negative $v_x$ for $x < 0.25 L_x$; $\left<\gamma\right> \gg \left<\gamma''\right>$ due to Lorentz transformation), also characterised by moderate particle density and intrinsic temperature $\left<\gamma''\right> / \Theta$, specific electric current $|j_z| / j_{\rm max}$ decreasing systematically with distance, very weak intrinsic electric field $E_z'$, and weak synchrotron emission. This is a structure that has all characteristics of minijets -- regular reconnection outflows of relativistic bulk velocity \citep{2009MNRAS.395L..29G,2011MNRAS.413..333N}.
We can easily recognise the conical geometry of the outflow region with parallel outflow velocity field (very low values of $v_y$), and oblique magnetic field lines crossing the outflow boundaries, as has been described by an analytical model of relativistic Petschek-type reconnection by \cite{2005MNRAS.358..113L}.
There is one qualitative difference from that model -- the magnetic field lines in the outflow region are not vertical, the magnetic field gradient $\partial B_x/\partial y$ does not vanish in that region and it is supported by the non-zero electric current density $j_z$.
We note that there is a roughly uniform vertical inflow of background plasma into the minijet region with velocity $v_y$ (reconnection rate) of the same order as that of the inflow into the main current layer.

To the right of the main current layer, we find a group of several plasmoids of various sizes, all propagating to the right at different velocities $v_x > 0$. The largest plasmoid can be seen centred at $x \simeq 0.725 L_x$, it is clearly slower than its smaller neighbour centred at $x \simeq 0.6 L_x$. The smaller plasmoid is in the process of merging with the large one, even though they both propagate in the same direction. This is a natural consequence of the inverse relation between the growth and bulk acceleration of plasmoids that has been first noticed by \cite{2016MNRAS.462...48S}.

%\clearpage
\subsection{Spacetime diagrams}
\label{sec_res_xtmap}

\begin{figure*}
\includegraphics[width=\textwidth]{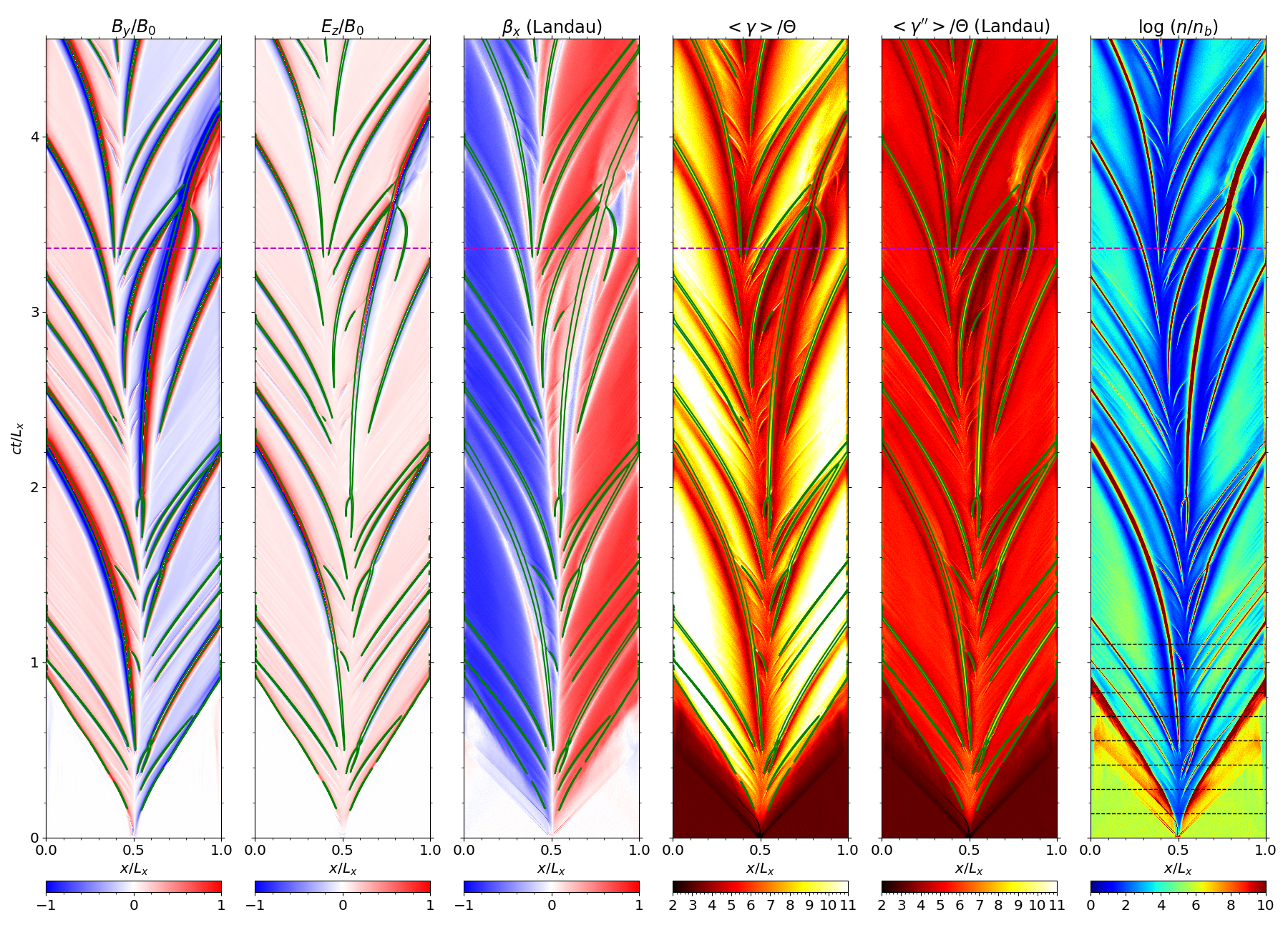}
\caption{Spacetime diagrams extracted along the reconnection midplane for the simulation {\tt s10Tm}.
From left to right we plot: (1) the magnetic field component $B_y /B_0$, (2) the electric field component $E_z / B_0$, (3) the Landau-type bulk velocity component $\beta_x$, (4) logarithm of the cold magnetisation parameter $\log(\sigma/\sigma_0)$, (5) mean particle Lorentz factor $\left<\gamma''\right> / \Theta$ measured in the Landau frame, and (6) logarithm of the particle number density $\log(n / n_{\rm b})$.
In the first five panels, we show the particle number density contours $n = 7n_{\rm b}$ with green solid lines. We indicate the simulation time $ct = 3.36 L_x$ corresponding to the simulation state shown in Figures \ref{fig:xprof} and \ref{fig:xymaps} with the magenta dashed horizontal lines. The black dotted horizontal lines in the last panel indicate the initial simulation states shown in Figure \ref{fig:xymap_init}.}
\label{fig:xtmaps}
\end{figure*}

Most of the information on evolution of current layers, and especially on the plasmoids, is contained along the reconnection midplane, this information can be presented very efficiently in the form of spacetime diagrams.
Following the practice of \cite{2015ApJ...815..101N}, the one-dimensional parameter $x$-profiles described in Section \ref{sec_meth_plasm} are combined into spacetime diagrams of high time resolution.
Figure \ref{fig:xtmaps} shows spacetime diagrams for several parameters for our reference simulation {\tt s10Tm}.

The spacetime diagrams reveal a sustained bifurcating outflow along the $x$ coordinate (regions of negative and positive velocity component $\beta_x$) and a variety of plasmoids (indicated by enhanced plasma density and sharp positive gradients of magnetic field component ${\rm d}B_y/{\rm d}x > 0$), most of which are generated along the $\beta_x \simeq 0$ line.
There are a few large plasmoids that evolve very slowly, their bulk velocities are non-relativistic, acceleration time scale is long, and hence they spend more than the light-crossing time $L_x/c$ before leaving the simulation domain.
On the other hand, there are many small plasmoids that accelerate quickly to relativistic bulk velocities, and they spend less than $L_x/c$ in the simulation domain.
Small plasmoids may either escape through the boundaries, or merge with a larger plasmoid.
As the velocity field is generally divergent, plasmoid mergers are relatively rare, typically they involve plasmoids of different sizes moving in the same direction (tail-on mergers).
A large plasmoid can attract nearby small plasmoids, and can even reverse their motion, e.g., Figure \ref{fig:xymaps} shows that at $ct/L_x = 3.36$ a small plasmoid centred at $x \simeq 0.85 L_x$ has $\beta_x > 0$, however, the spacetime diagrams reveal that it will merge with the large plasmoid located on its left side by $ct/L_x \simeq 3.6$.

The regions between plasmoids are characterised by roughly uniform electric field component $E_z \simeq 0.1 B_0$ and by two-value magnetic field component $B_y \simeq \pm 0.15 B_0$ (positive where $\beta_x < 0$ and negative where $\beta_x > 0$). The uniformity of electric field measured in the simulation frame is consistent with the uniform reconnection rate indicated in the $\beta_y$ panel of Figure \ref{fig:xymaps}. It is remarkable given the non-uniformity of electric field measured in local Eckart frames $E_z'$, as shown in another panel of Figure \ref{fig:xymaps}. This indicates a smooth connection between the minijet outflows (where $E_z'$ is close to zero) and the proper magnetic diffusion areas (where $E_z'$ is strong).

High density regions of the plasmoids -- essentially the plasmoid cores -- are initially characterised by enhanced temperatures (mean particle energies measured in the local Landau frames reaching values of $\left<\gamma''\right> \simeq 10\Theta$) due to the heating of particles in the reconnection process.
Our spacetime diagram of $\left<\gamma''\right> / \Theta$ demonstrates that the cores of large plasmoids cool down gradually.

\begin{figure*}
\includegraphics[width=\textwidth]{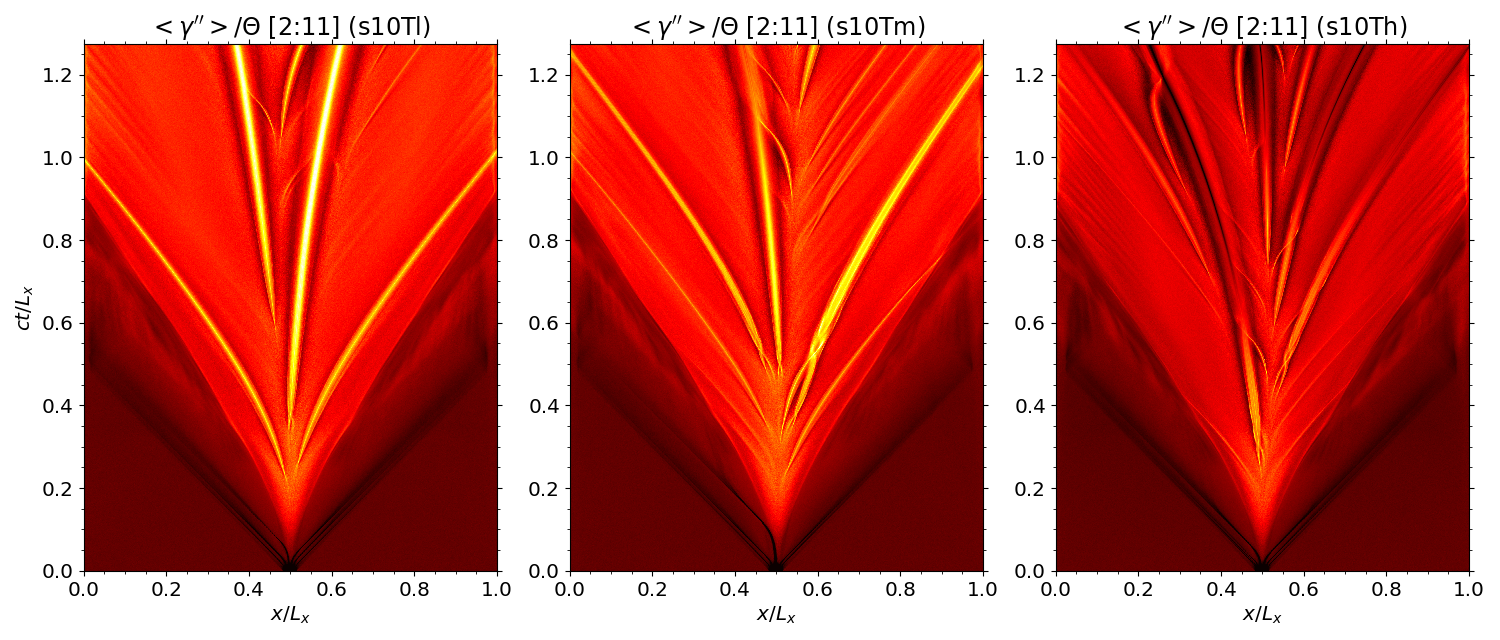}
\caption{Spacetime diagrams of average particle Lorentz factor $\left<\gamma''\right> / \Theta$ measured in the Landau frame compared for three simulations characterised by different radiative cooling efficiencies.}
\label{fig:xtmaps_temp}
\end{figure*}

In Figure \ref{fig:xtmaps_temp}, we compare the spacetime diagrams of $\left<\gamma''\right> / \Theta$ for three simulations with different nominal cooling lengths $l_{\rm cool}$. In the simulation {\tt s10Tl}, in which $l_{\rm cool} > L_x$, we find that plasmoid temperatures are the highest, showing no signs of cooling over the light-crossing time scale $L_x/c$. On the other hand, in the simulation {\tt s10Th}, the plasmoid temperatures are the lowest, they eventually become even lower than for the plasma between the plasmoids.

\begin{figure*}
\includegraphics[width=\columnwidth]{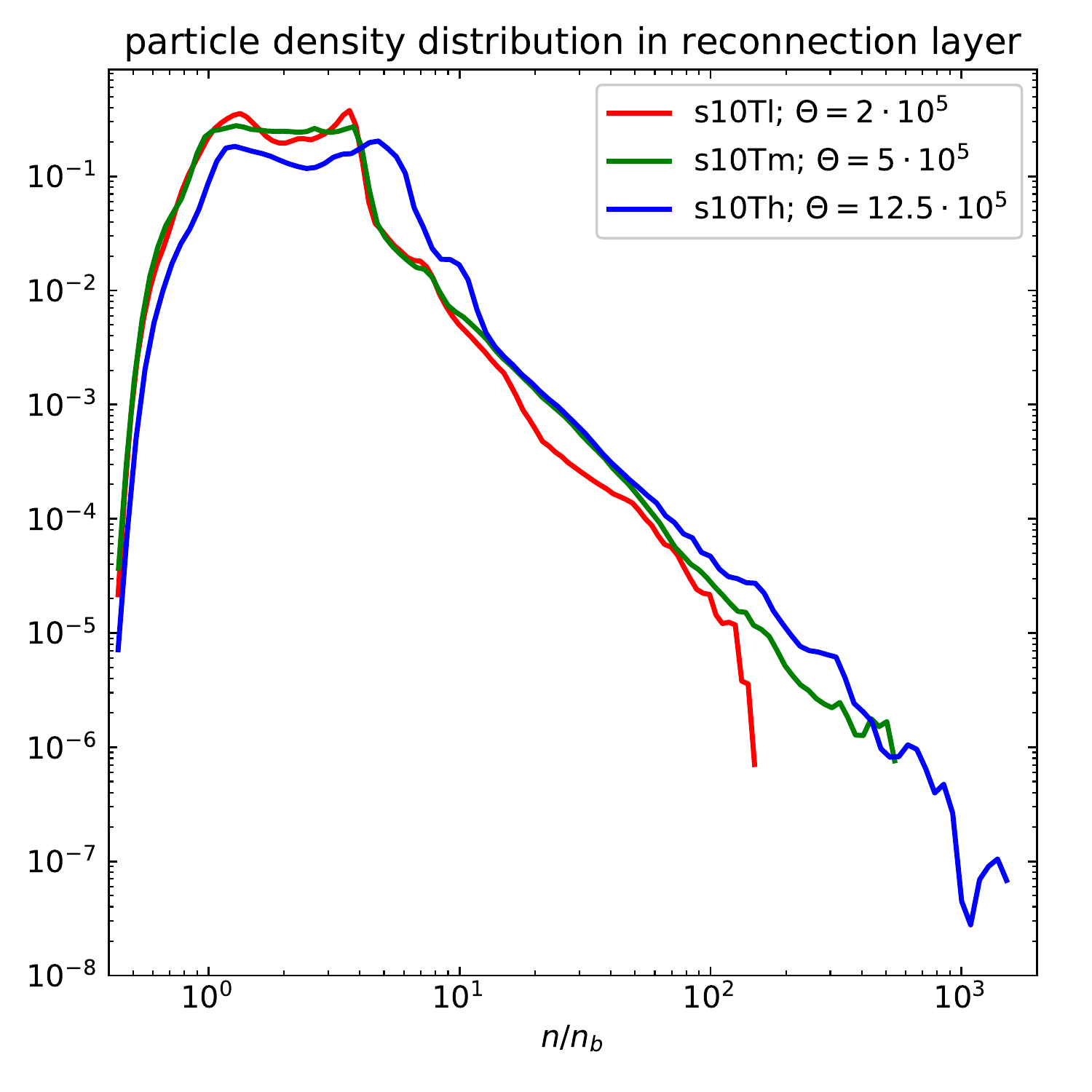}
\includegraphics[width=\columnwidth]{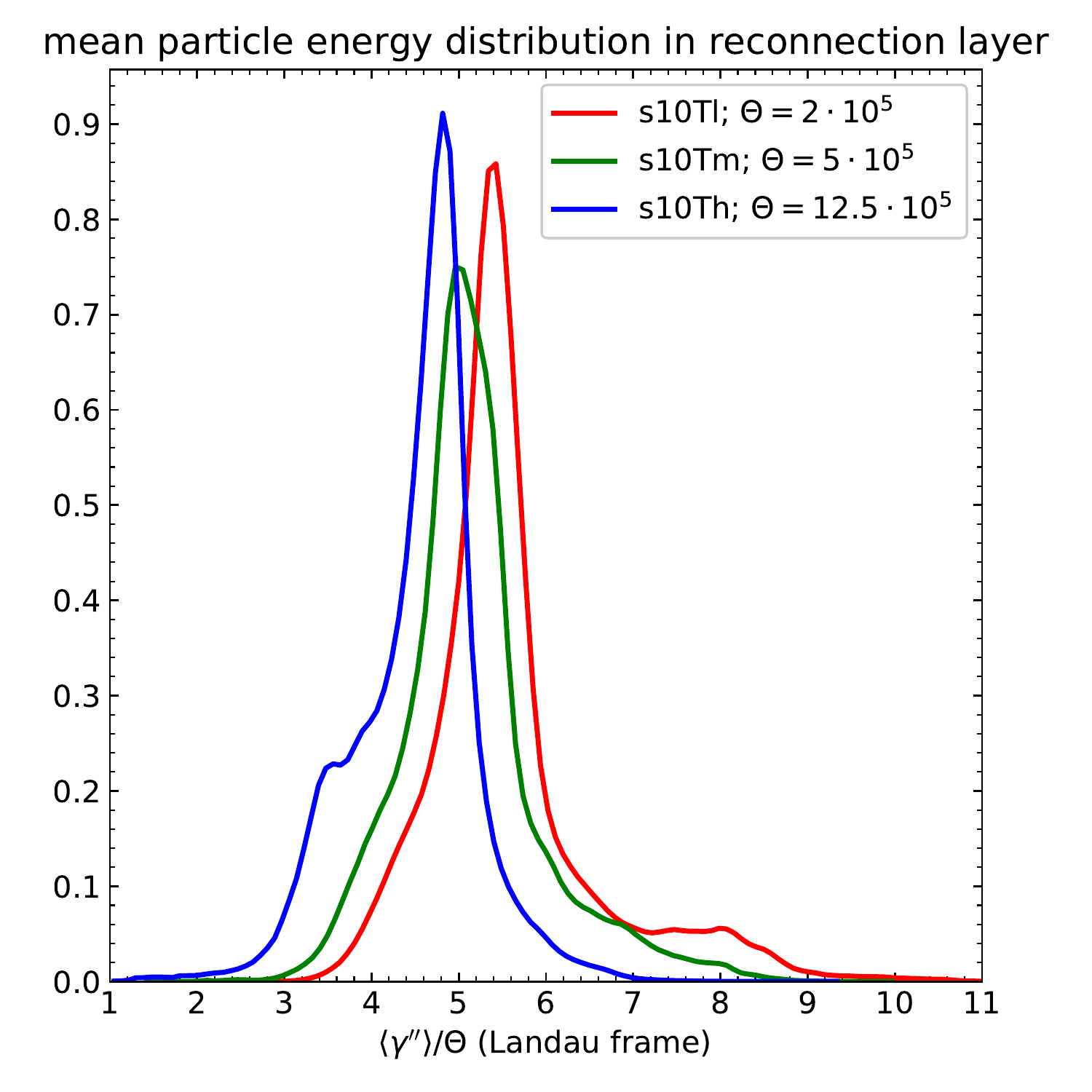}
\caption{Distributions of particle density $n / n_{\rm b}$ (left panel) and mean particle energy $\left<\gamma''\right> / \Theta$ calculated in the Landau co-moving frame (right panel) extracted from the spacetime diagrams probing narrow regions along the reconnection layers for the 3 simulations with $\sigma_0 = 10$ that differ by initial gas temperature $\Theta$, and hence by the radiative cooling efficiency (higher $\Theta$ corresponds to more efficient cooling).}
\label{fig:hist_dens_temp}
\end{figure*}

\subsection{Plasma density and temperature distributions}
\label{sec_res_dens_temp}

The left panel of Figure \ref{fig:hist_dens_temp} compares the distributions of particle number density $n / n_{\rm b}$ based on the spacetime diagrams of 3 simulations with $\sigma_0 = 10$ and with different efficiencies of radiative cooling.
The distributions form power-law tails with the slope of $\sim -2.3$, independent of the cooling efficiency.
However, the highest values of the particle density increase with stronger cooling, from $n_{\rm max} \simeq 150 n_{\rm b}$ for the simulation {\tt s10Tl} to $n_{\rm max} \simeq 1500 n_{\rm b}$ for the simulation {\tt s10Th}.
The highest particle densities are realised in the cores of large plasmoids.
Strong radiative cooling removes the gas pressure support, allowing the plasmoid cores to contract further.

The right panel of Figure \ref{fig:hist_dens_temp} compares the distributions of mean particle energy $\left<\gamma''\right> / \Theta$ measured in the local Landau frames, also extracted from the spacetime diagrams of simulations with $\sigma_0 = 10$.
This shows that the distributions are strongly peaked at values that decrease slightly with increasing cooling efficiency, from $\left<\gamma''\right>_{\rm peak} \simeq 5.4\Theta$ for the simulation {\tt s10Tl} to $\left<\gamma''\right>_{\rm peak} \simeq 4.8\Theta$ for the simulation {\tt s10Th}.

%\clearpage
\subsection{Individual plasmoids}
\label{sec_res_plasm}

\begin{figure*}
\includegraphics[width=\textwidth]{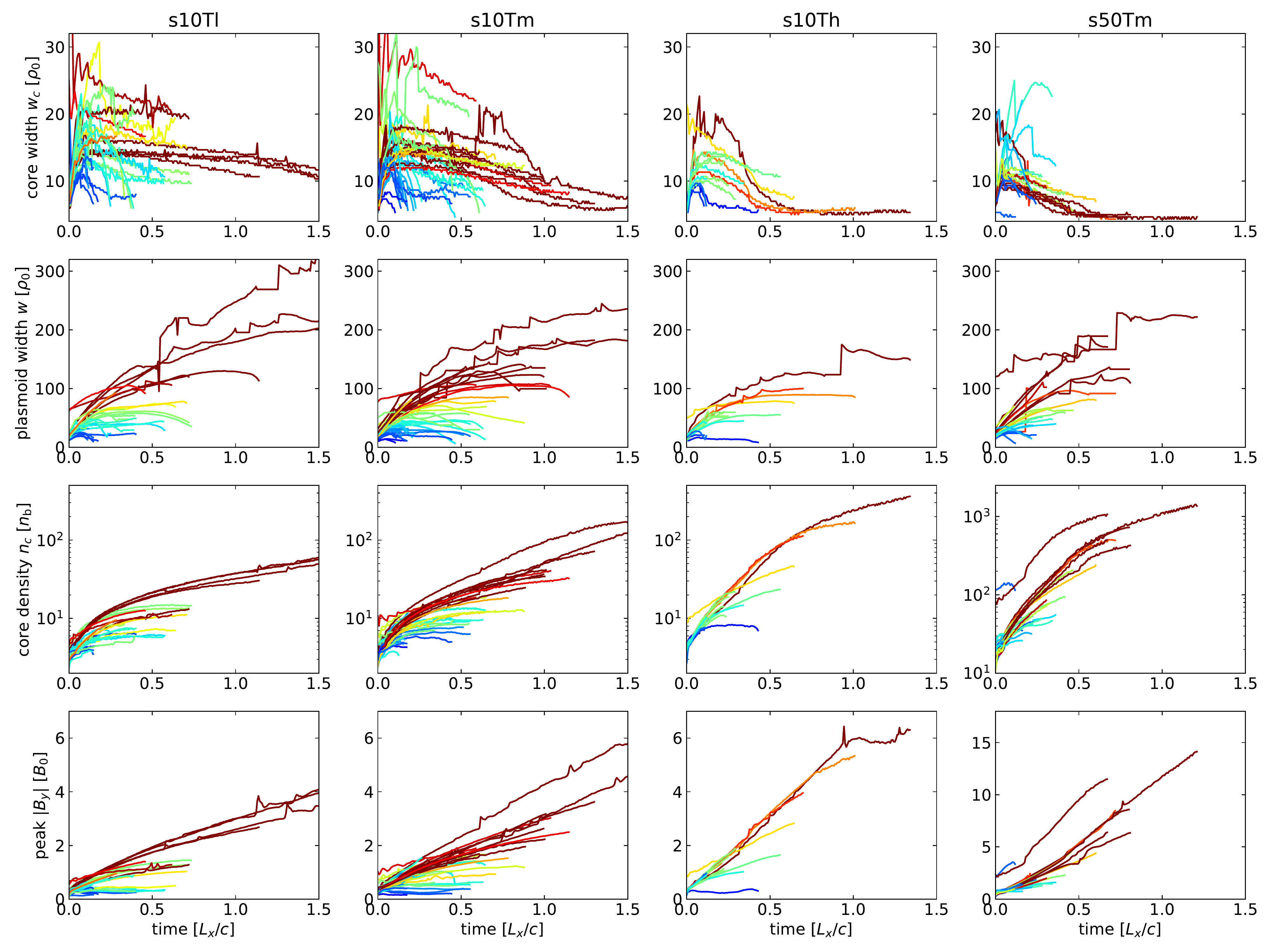}
\caption{The histories of individual plasmoids compared for the four large simulations.
From the top, the rows present:
(1) plasmoid core width $w_c / \rho_0$;
(2) total plasmoid width $w / \rho_0$;
(3) mean core density $n_c / n_b$;
(4) peak magnetic field strength $|B_y| / B_0$.
For each plasmoid, time is measured relative to its first appearance.
The line colour indicates the peak total plasmoid width.}
\label{fig:plasmoids_widths}
\end{figure*}

\begin{figure*}
\includegraphics[width=\textwidth]{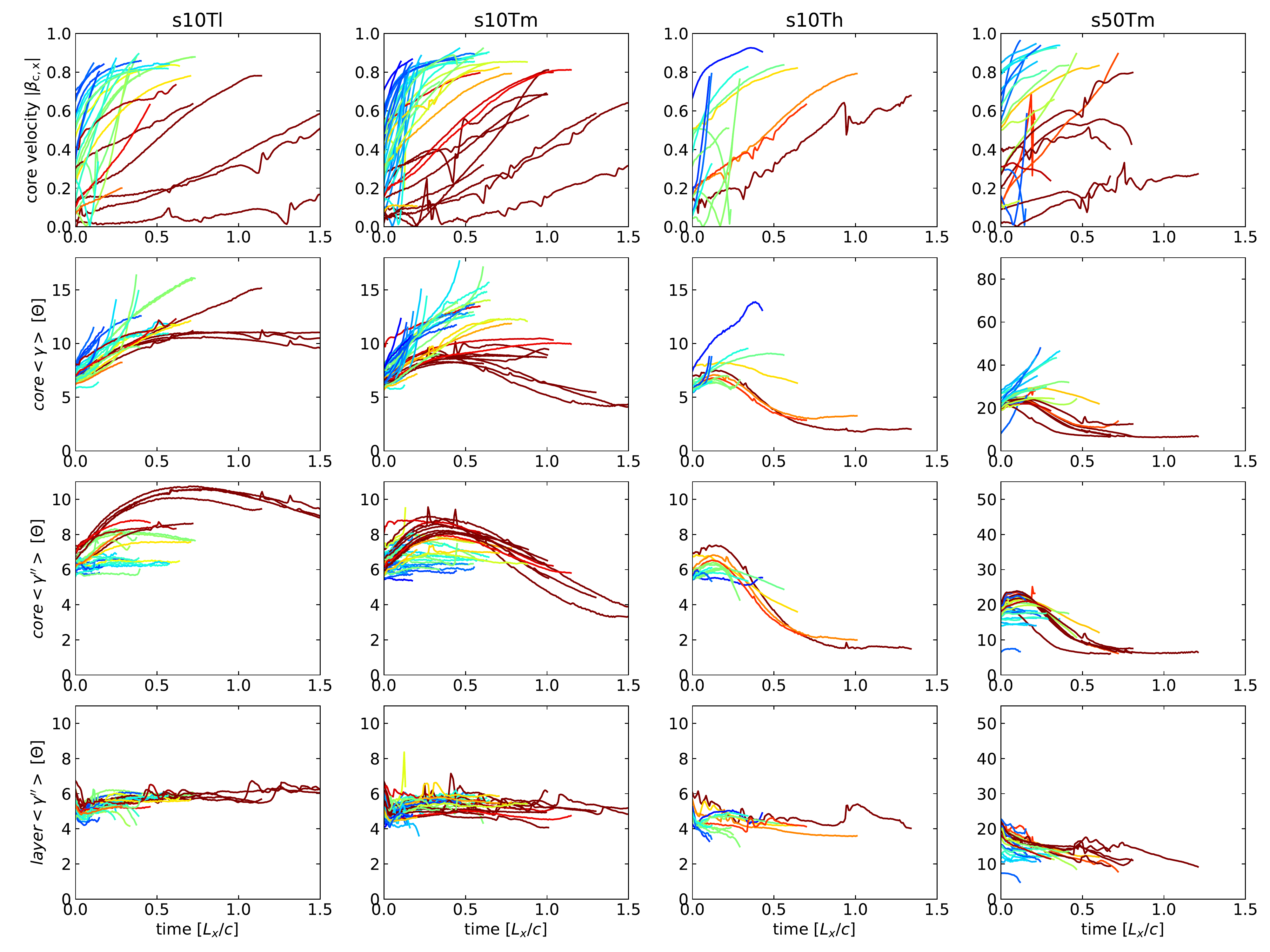}
\caption{Further histories of individual plasmoids.
From the top, the rows present:
(1) mean core velocity $|\beta_{\rm c,x}|$;
(2) mean particle energy $\left<\gamma\right>_c / \Theta$ measured in the simulation frame, averaged over the plasmoid core;
(3) mean Landau-frame particle energy $\left<\gamma''\right>_c / \Theta$ of the plasmoid core;
(4) mean Landau-frame particle energy $\left<\gamma''\right>_l / \Theta$ of the plasmoid layer.
See Figure \ref{fig:plasmoids_widths} for more description.}
\label{fig:plasmoids_temperatures}
\end{figure*}

\begin{figure*}
\includegraphics[width=\textwidth]{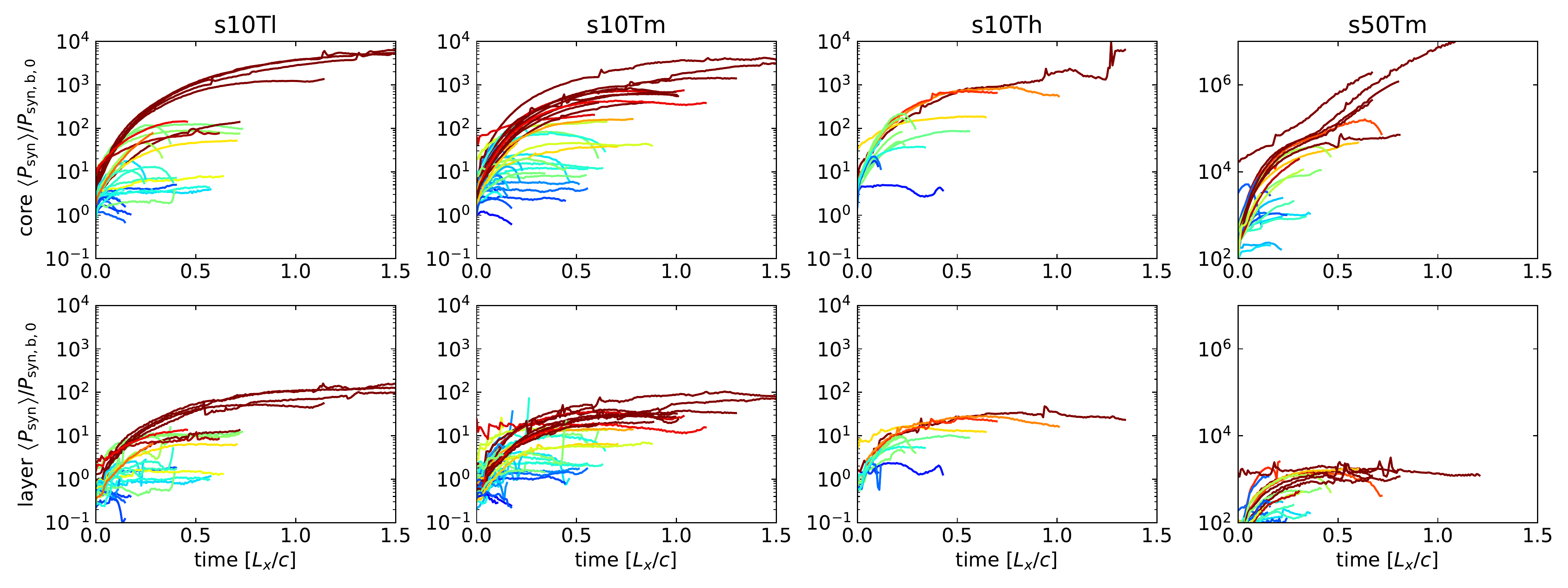}
\caption{Further histories of individual plasmoids.
From the top, the rows present:
(1) mean synchrotron emissivity $\left<\mathcal{P}_{\rm syn,c}\right> / \mathcal{P}_{\rm syn,b,0}$ of the plasmoid core,
(2) mean synchrotron emissivity $\left<\mathcal{P}_{\rm syn,l}\right> / \mathcal{P}_{\rm syn,b,0}$ of the plasmoid layer.
See Figure \ref{fig:plasmoids_widths} for more description, and Eq. (\ref{eq_Psyn0}) for the definition of $\mathcal{P}_{\rm syn,b,0}$.}
\label{fig:plasmoids_Esyn}
\end{figure*}

Figures \ref{fig:plasmoids_widths}, \ref{fig:plasmoids_temperatures} and \ref{fig:plasmoids_Esyn} present the histories of individual plasmoids compared for the four simulations listed in Table~\ref{tab:param}.
The plasmoid histories are compiled from the parameters determined independently for each timestep in automated plasmoid identification algorithm described in Section \ref{sec_meth_plasm}. As such, they are subject to some irregularities, in particular due to shifting positions of the minima of magnetic potential $A_z$ (that define the plasmoid boundaries), especially for large plasmoids merging with the small ones. In order to reduce these irregularities, we chose not to record plasmoid histories in the following circumstances: (1) when there is no gap between two separate plasmoids (i.e., they share the same local $A_z$ minimum); (2) when the plasmoid layer boundary is less than $100\;{\rm d}x$ from the left/right domain boundary.

We present separately the total plasmoid widths $w$ (dominated by the widths of plasmoid layers) and the widths of their cores $w_c$, noting that there is no correlation between them.
In all simulations, the plasmoid core widths, after a brief formation phase, show a decreasing trend.
On the other hand, the total plasmoid widths grow systematically in time for every simulation.
%The widths of large plasmoids show occasional dips that correspond to their mergers with smaller plasmoids, during which the positions of the minima of magnetic potential $A_z$ (that define the plasmoid boundaries) shift temporarily inwards.

We find that the densities of plasmoid cores $n_c$ grow systematically in time for all simulations.
The larger the plasmoid, the denser its core becomes.
For simulations with $\sigma_0 = 10$, higher core densities are reached for higher radiative efficiencies.
In the case of $\sigma_0 = 50$, the core densities are even higher, roughly in proportion to $\sigma_0$.

The histories of plasmoid core velocities $\beta_{c,x}$ confirm the picture discussed before of small plasmoids being accelerated very rapidly to relativistic velocities and of large plasmoids being accelerated slowly only to mildly relativistic velocities.

The mean particle energies of plasmoid cores, measured in the simulation frame, are somewhat higher for small plasmoids. However, when measured in the Landau frame of the core, they are instead higher for large plasmoids until the radiative cooling effects become significant. The difference is mainly due to relativistic bulk motions of the small plasmoids.
In particular, in the simulations {\tt s10Tl} and {\tt s10Tm} characterised by weak/moderate cooling, small plasmoids have $\left<\gamma''\right>_c \simeq 6\Theta$, roughly constant in time, while large plasmoids can reach $\left<\gamma''\right>_c \simeq 11\Theta$ in the case {\tt s10Tl}.
In the simulation {\tt s10Th} characterised by strong cooling, the intrinsic mean particle energies of the core are reduced down to $\left<\gamma''\right>_c \simeq 2\Theta$ by $L_x/c$.
In the simulation {\tt s50Tm}, the cores of large plasmoids are heated up to $\left<\gamma''\right>_c \simeq 23\Theta$ before cooling down to $\left<\gamma''\right>_c \lesssim 10\Theta$.

In contrast, the plasmoid layers are characterised by similar and stable intrinsic temperatures $\left<\gamma''\right>_c \simeq (4-6)\Theta$ for all simulations with $\sigma_0 = 10$ (and $\left<\gamma''\right>_c \sim 15\Theta$ for $\sigma_0 = 50$), showing virtually no signs of radiative cooling.

Most of the plasmoids show their synchrotron emissivity building up in time, both for the core and for the layer.
Some of the smaller plasmoids show a very slow decline.
%some plasmoids show a rapid decline towards the end of their histories.
The cores of large plasmoids produce significantly more (2-3 orders of magnitude) synchrotron emission per volume element than the cores of small plasmoids, and also about 2 orders of magnitude higher emission density than the large plasmoid layers.
Our basic finding is that synchrotron emission of the plasmoids is sustained over a long term, irrespective of the cooling efficiency.
For the cores of large plasmoids that undergo the most efficient radiative cooling in our high-temperature simulations, it appears that systematically increasing core density, as well as systematically increasing magnetic field strength are able to offset the reduction in particle mean energy.
In the case of $\sigma_0 = 50$, we find a systematic increase of the emission from the cores of large plasmoids to the levels up to $10^7 \mathcal{P}_{\rm syn,b,0}$.

%\clearpage
\subsection{Individual particles}
\label{sec_res_part}

\begin{figure}
\includegraphics[width=\columnwidth]{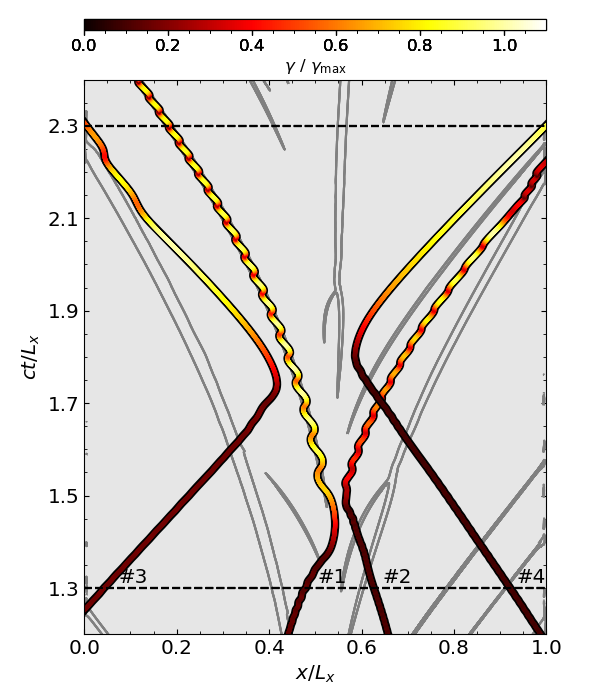}
\caption{Spacetime diagram of the tracks of selected energetic particles, the acceleration of which is characterised in detail in Figure \ref{fig:Particle_track}. The line colour indicates the instantaneous particle energy measured in the simulation frame. Particle density contours $n = 7n_{\rm b}$ are indicated with grey lines.}
\label{fig:Particle_xt_track}
\end{figure}

\begin{figure*}
\includegraphics[width=0.495\textwidth]{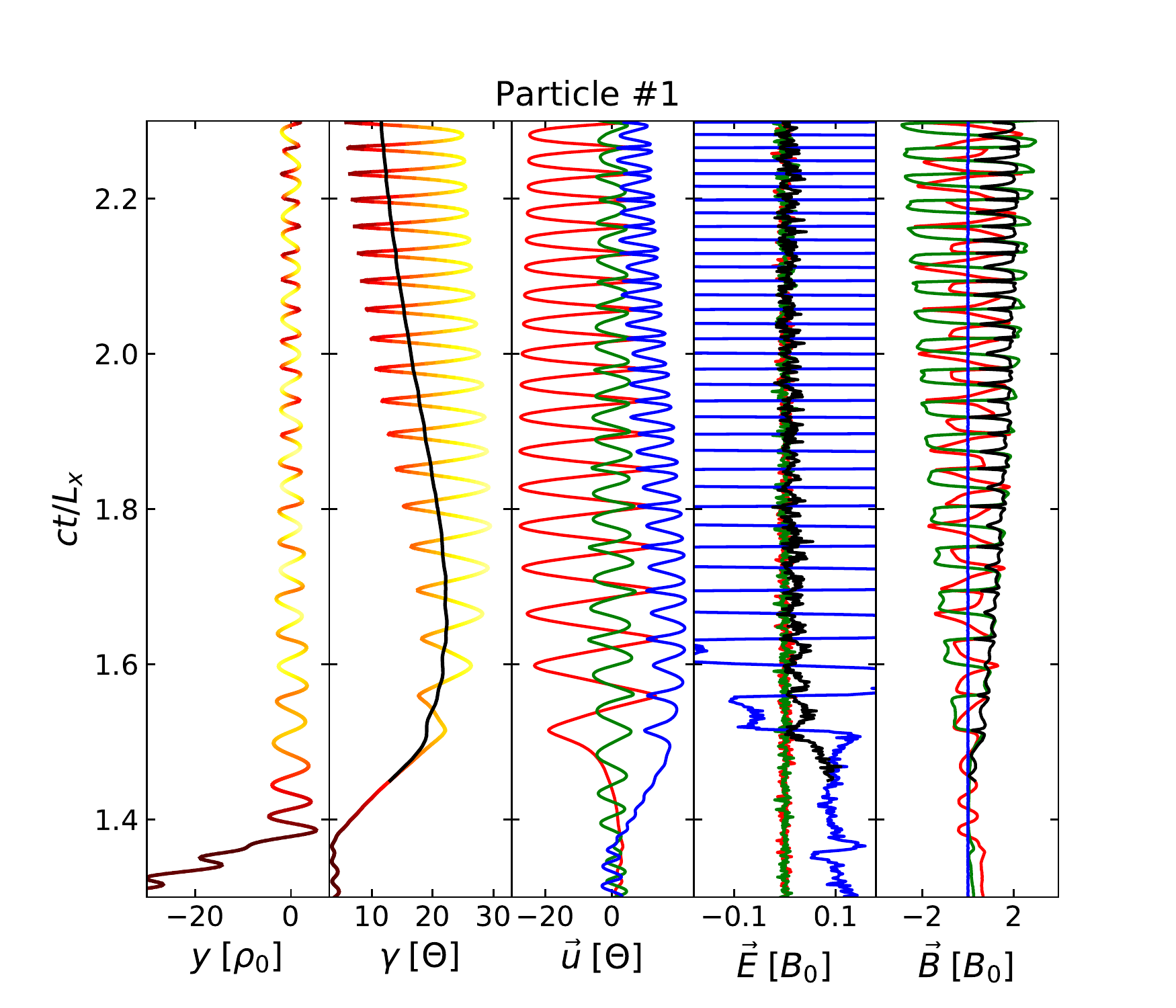}
\includegraphics[width=0.495\textwidth]{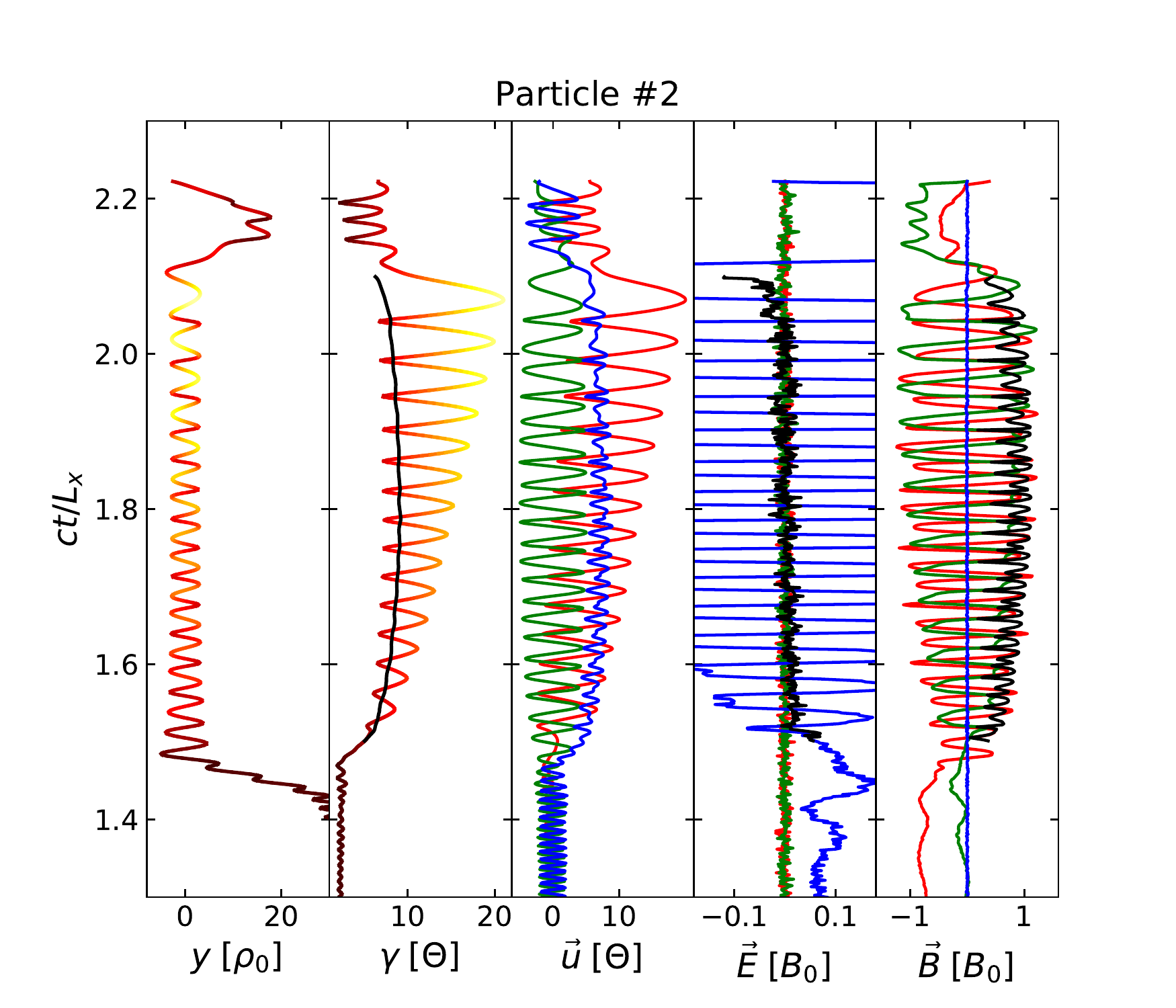}
\includegraphics[width=0.495\textwidth]{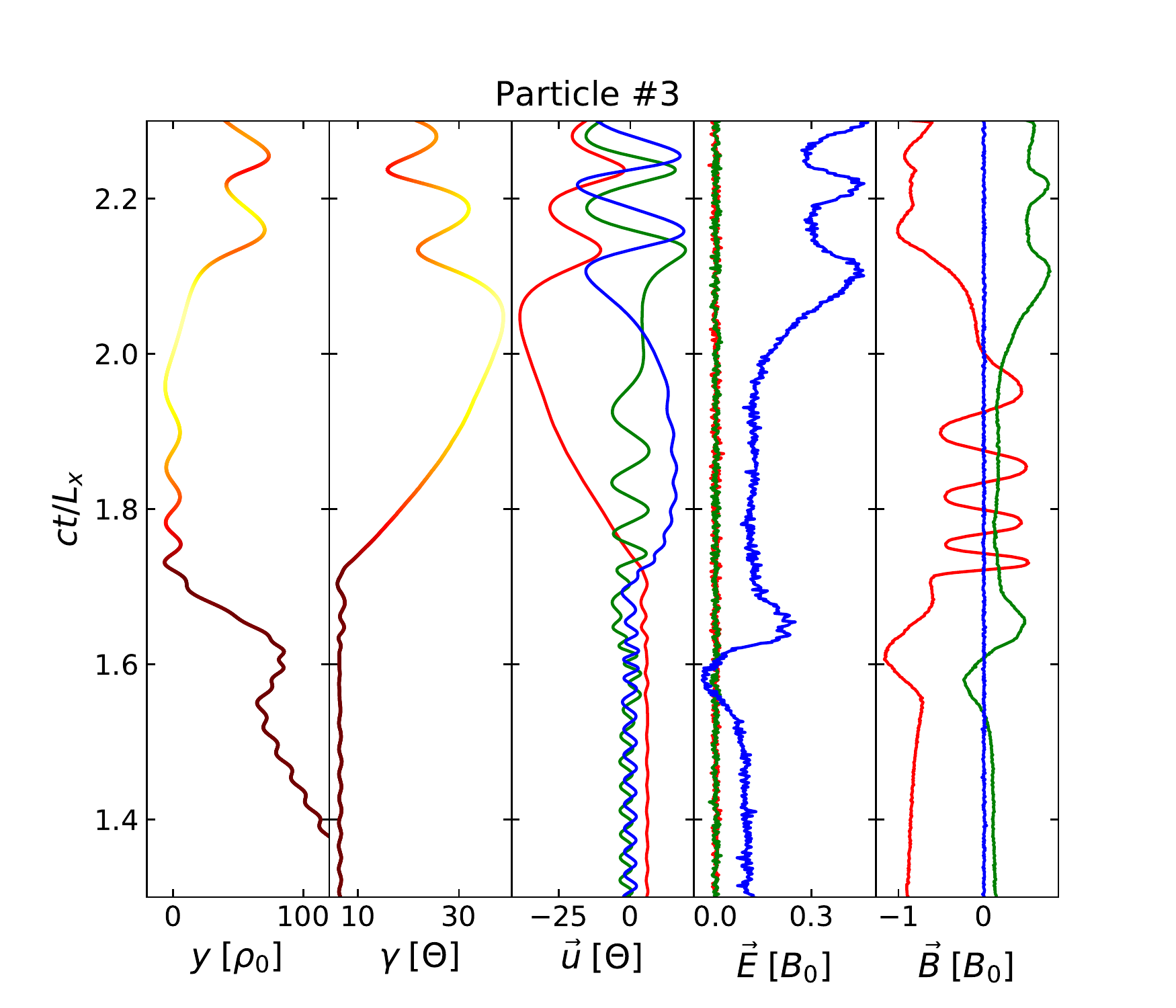}
\includegraphics[width=0.495\textwidth]{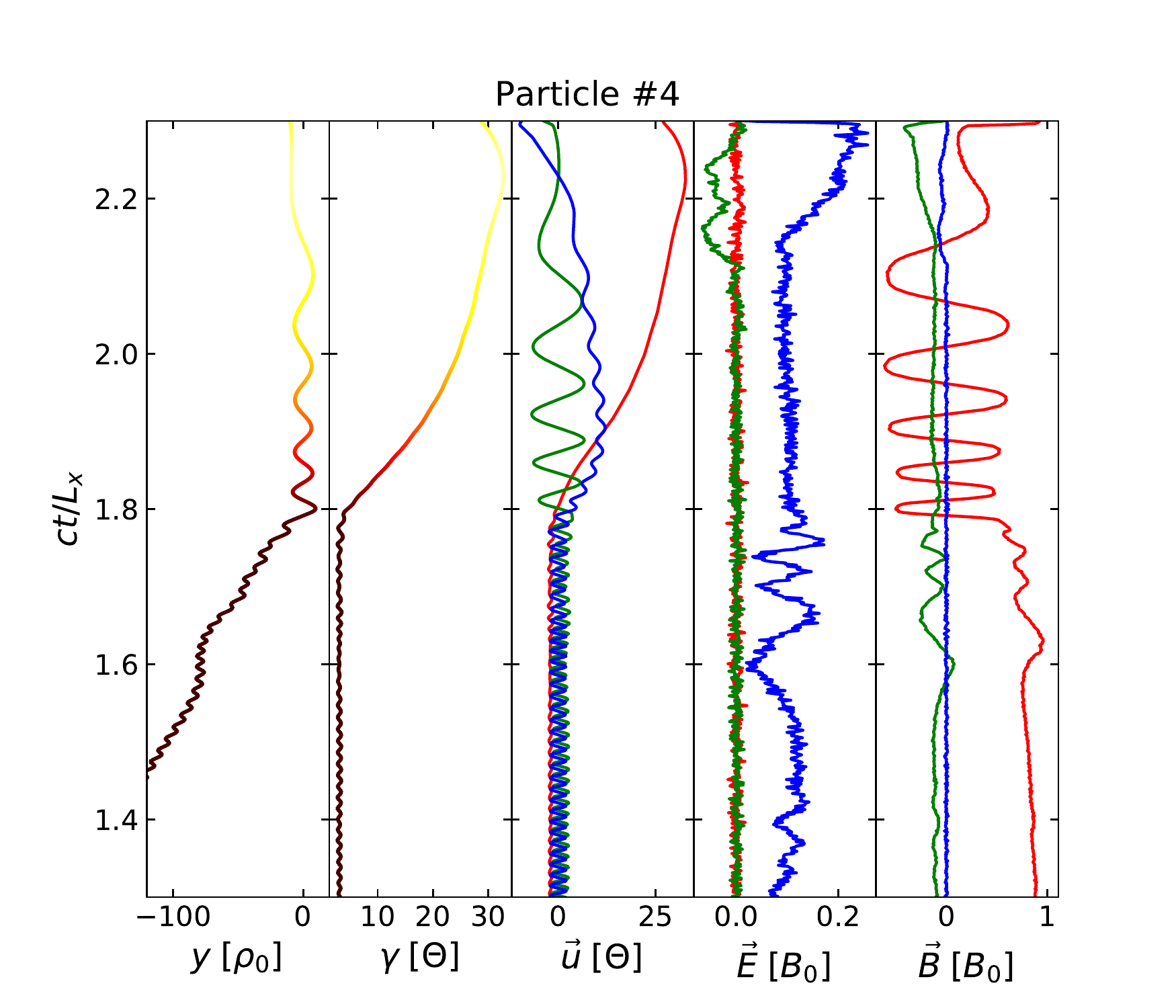}
\caption{Acceleration histories for selected energetic particles, the $x$-positions of which are shown in Figure \ref{fig:Particle_xt_track}. For each particle we present detailed information as functions of simulation time on five panels, from the left: (1) the $y$-position measured from the reconnection midplane, (2) particle Lorentz factor $\gamma$ (also indicated with a colour scale on panels 1 and 2) normalised to $\Theta$, (3) three components of the particle 4-velocity $\bm{u}$ ($x$ - red, $y$ - green, $z$ - blue) normalised to $\Theta$, (4) three components of the local electric field $\bm{E}$ in units of $B_0$, (5) three components of the local magnetic field $\bm{B}$ in units of $B_0$. For particles \#1 and \#2, the black lines indicate parameter values (Lorentz factor $\gamma'$, electric field $E_\parallel'$ and magnetic field $B_\perp'$) measured in the co-moving frame of a plasmoid to which the particle is attached.}
\label{fig:Particle_track}
\end{figure*}

Here we describe the behaviour of four selected energetic positrons (denoted as particles \#1 --- \#4), for which a detailed history has been recorded. The spacetime tracks of these particles are indicated in Figure \ref{fig:Particle_xt_track}. From this one can see that particles \#1 and \#2 become trapped in different plasmoids, while particles \#3 and \#4 become accelerated in the low-density regions between plasmoids.

Figure \ref{fig:Particle_track} presents the detailed acceleration histories of these four particles, all in the same time window. Particle \#1 becomes accelerated to Lorentz factor $\gamma \simeq 22\Theta$ in a single episode. The beginning of acceleration episode coincides with the particle becoming trapped in the reconnection midplane ($|y| < 5\rho_0$). We can also see that the acceleration episode coincides with a formation of a new plasmoid that traps the particle also in the $x$ coordinate. The particle experiences mainly the electric field component $E_z \sim 0.1B_0$, and its momentum gain is at first mainly in the $z$ direction, later also in the $x$ direction. After the acceleration episode, the particle oscillates around the plasmoid centre, which results in oscillations of its Lorentz factor $\gamma$ measured in the simulation frame. However, its Lorentz factor $\gamma'$ measured in the plasmoid frame shows a slow gradual decline in time, which we attribute to the radiative cooling.

Particle \#2 shows a similar behaviour to particle \#1, the acceleration episode also coincides with the formation of a new plasmoid. In this case, the acceleration episode is shorter and the energy gain is also lower, with acceleration proceeding in similar electric fields. The radiative cooling is less efficient, and it should be noted that the co-moving perpendicular magnetic field is about twice weaker. At $ct \simeq 2.1L_x$, the plasmoid in which particle \#2 is trapped merges into a larger plasmoid on its right side. We see that this results in particle \#2 losing most of its energy measured in the simulation frame. It appears that this particle did not experience direct deceleration by strong electric field, instead it just happened to be at its lower energy level in the oscillation cycle at the moment of plasmoids merger. It has also been kicked out from the reconnection midplane, settling briefly at $y \sim 15\rho_0$, trailing behind the merged plasmoid before it exits the right boundary of the domain.

Particle \#3 was accelerated in a long acceleration episode under constant electric field $E_z \sim 0.12B_0$ after becoming trapped in the $y$ coordinate to the reconnection midplane. The $y$ and $B_x$ data reveal a typical Speiser orbit with gradually increasing period. The acceleration episode is interrupted at $ct \simeq 2.1L_x$, when the particle begins to interact with a large plasmoid, which forces the particle away from the reconnection midplane. The particle bounces twice off the trailing edge of the plasmoid before it exits the left boundary of the domain. Particle \#4 shows another example of Speiser-orbit acceleration in the low-density region of reconnection midplane that is enabled by trapping the particle in the $y$ coordinate.

%\clearpage
\subsection{Synchrotron emissivity and lightcurves}
\label{sec_res_lc}

\begin{figure}
\includegraphics[width=\columnwidth]{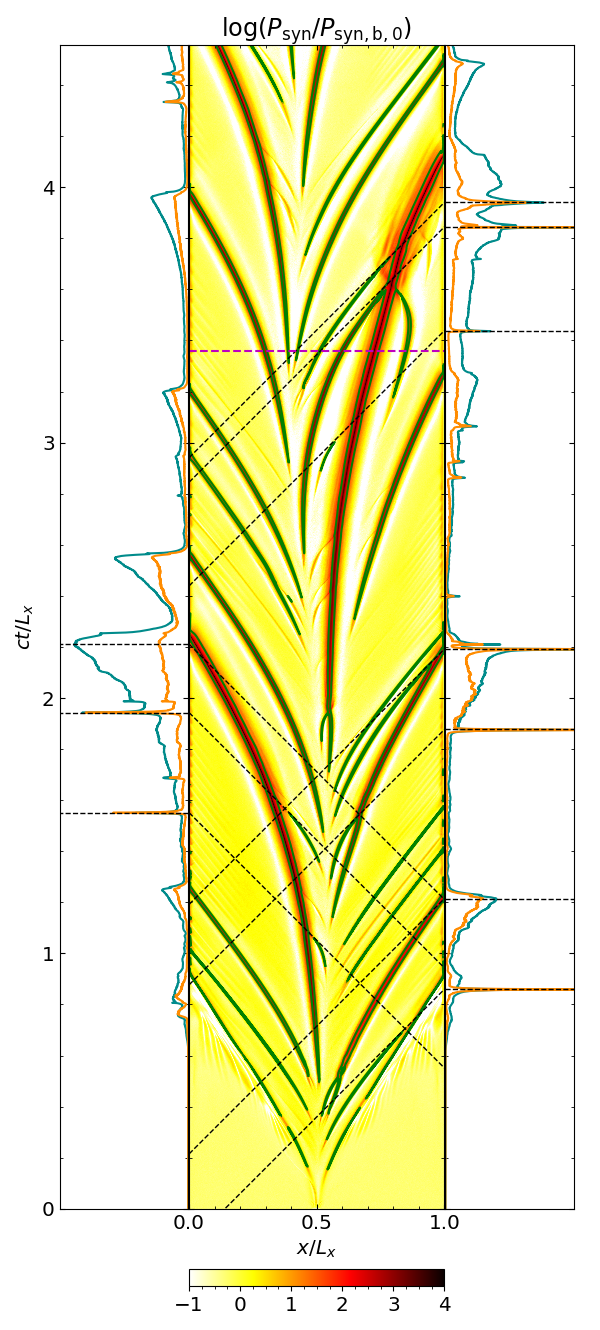}
\caption{The left and right panels show synchrotron lightcurves measured by observers placed at left and right sides of the reconnecting layer, respectively. The cyan and orange lines correspond to the two selected frequency bands indicated in Figure \ref{fig:spec_syn}.
The middle panel shows the spacetime distribution of total synchrotron power normalised to the value $\mathcal{P}_{\rm syn,b,0}$ defined in Eq. (\ref{eq_Psyn0}).
The particle number density contours $n = 7n_{\rm b}$ are shown with green solid lines.
The black dashed lines represent the lightcones corresponding to selected features in either light curve.
The magenta dashed line indicates the simulation time $ct = 3.36 L_x$ corresponding to the simulation state shown in Figures \ref{fig:xprof} and \ref{fig:xymaps}.}
\label{fig:lightcurves_xtmap_Esyn}
\end{figure}

Figure \ref{fig:lightcurves_xtmap_Esyn} presents the spacetime distribution of the total synchrotron power emitted from the reconnection midplane in all directions for simulation {\tt s10Tm}, and lightcurves that would be received by two distant observers placed at the opposite sides of the simulation domain, one on the left ($-x$ axis), one on the right ($+x$ axis).
While the spacetime diagram of synchrotron power is based on the $x$-profiles integrated over narrow stripes of $|y| < \delta/2$, the lightcurves are calculated from a much wider region of $|y| < L_x/2$ to ensure contribution from the whole plasmoids and their broader surroundings.

The synchrotron emission is strongly concentrated along the plasmoid trajectories. As we have noted in Section \ref{sec_res_plasm}, large plasmoids produce significantly more synchrotron emission. The emissivity of the large plasmoid cores exceeds the emissivity of the background plasma by four orders of magnitude.

Lightcurves received by either observer are completely different, indicating a high level of anisotropy.
The most conspicuous features of the lightcurves are very sharp and bright flares.
We can identify the origin of these flares by drawing the corresponding lightcones on the spacetime diagram to the events of small plasmoids approaching the observer with relativistic velocity and merging with a large target plasmoid that is also approaching the observer.
Zooming up on these flares, we find their characteristic time scales of $\tau \sim \rho_0/c$, so they are basically unresolved\footnote{Contributions to every lightcurve are recorded at every simulation timestep at the temporal resolution of ${\rm d}t$.}.
Because of their very short duration, the contribution of these flares to the overall radiation fluence is rather insignificant.
The lightcurves also contain smooth structures characterised by relatively long rise and short decline. The can be attributed to large plasmoids propagating with mildly relativistic velocities, and the sharp declines observed in the lightcurves coincide with these plasmoids exiting the simulation domain. The contributions of these plasmoids to the lightcurves is not complete, since their emission is not contained within the simulation boundaries, as we have already noted in Section \ref{sec_res_plasm}.

Comparing the lightcurves recorded in two frequency bands\footnote{The lightcurves are presented in the same arbitrary units equivalent to the $\nu F_\nu$ flux density.}, we find that in general the contribution from large plasmoids is higher in the lower-frequency band (cyan lines), while the flares produced by small plasmoids are stronger in the higher-frequency band (orange lines).
Lightcurves obtained from different simulations are qualitatively very similar, in particular the level of radiative cooling efficiency does not clearly affect the lightcurve characteristics.

\subsection{Energy conservation}
\label{sec_res_enecons}

\begin{figure}
\includegraphics[width=\columnwidth]{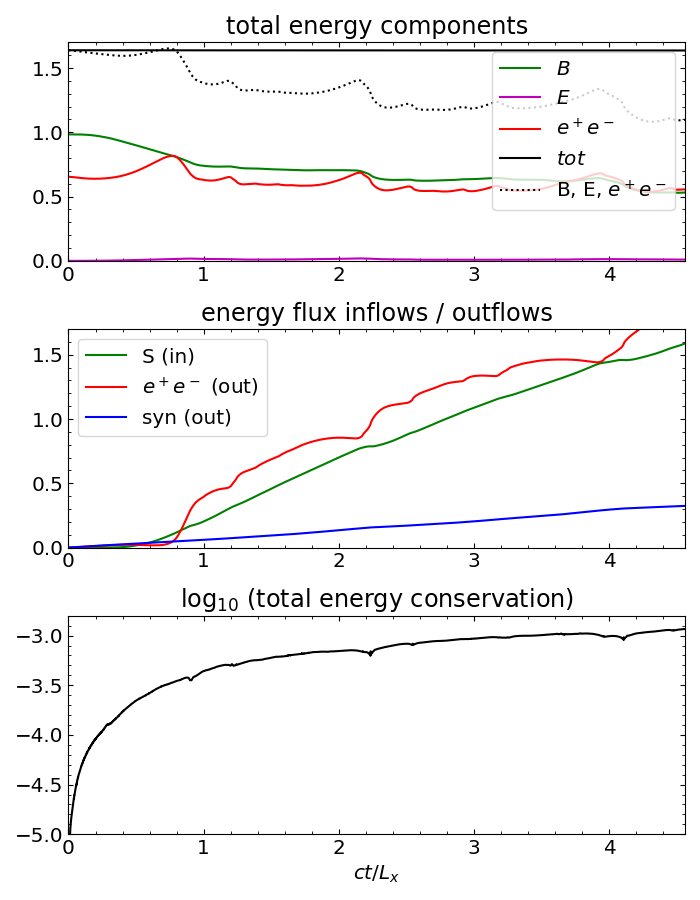}
%DOUBLE PRECISION, PARAMETER, PUBLIC :: x1_reg = xmin + 2.0*d_abs*dx
%DOUBLE PRECISION, PARAMETER, PUBLIC :: x2_reg = xmax - 2.0*d_abs*dx
%DOUBLE PRECISION, PARAMETER, PUBLIC :: y1_reg = ymin + 0.5*(ymax-ymin) - 0.25*(xmax-xmin)
%DOUBLE PRECISION, PARAMETER, PUBLIC :: y2_reg = ymin + 0.5*(ymax-ymin) + 0.25*(xmax-xmin)
\caption{Top panel: mean energy densities, normalised to initial magnetic energy density \(U_{\rm B,0}\), calculated for the analysis region $\mathcal{R}$, defined by $2\Delta_{\rm abs} < x < L_x - 2\Delta_{\rm abs}$ and $-L_x/4 < y < L_x/4$, for the simulation {\tt s10Tm}.
We plot the contributions from magnetic fields (green), electric fields (magenta), all particles ($e^+$ and $e^-$; red), and total conserved energy including inflows and outflows (black).
The black dotted line shows the total instantaneous energy contained in $\mathcal{R}$, including contributions from the magnetic and electric fields, and all particles.
Middle panel: energy flux inflows and outflows: the inflowing Poynting flux (green), outflowing particle energy flux (red), and total synchrotron emission (blue).
Bottom panel: conservation of the total energy of region $\mathcal{R}$, including inflows and outflows.}
\label{fig:energy_conservation}
\end{figure}

Because of the use of open boundaries with steady injection of fresh particles, our simulations do not conserve energy globally. In order to evaluate the efficiency of energy transformations, we defined a fixed region $\mathcal{R}$ centred around the reconnection midplane between the left/right absorbing boundary layers, defined by $2\Delta_{\rm abs} < x < (L_x - 2\Delta_{\rm abs})$ and $-L_x/4 < y < L_x/4$.
In addition to the instantaneous energy contained in $\mathcal{R}$ in the form of magnetic and electric fields, as well as in the particles, we also calculate the cumulative energy emitted by all particles in the synchrotron process, and the fluxes of particles and electromagnetic fields (i.e., the Poynting flux) inflowing/outflowing across the $\mathcal{R}$ boundaries.

In Figure \ref{fig:energy_conservation}, we present the time evolution of different forms of energy contained in the region $\mathcal{R}$ for the simulation {\tt s10Tm}.
At the beginning of the simulation, the region $\mathcal{R}$ is dominated by magnetic energy ($\mathcal{E}_{\rm B,0} \simeq 0.6\,\mathcal{E}_{\rm tot}$).
The initial ($ct/L_x < 0.6$) energization of particles at the cost of magnetic energy is due to the trigger mechanism.
This is followed by the somewhat erratic variation of the particle energy, which reflects systematic heating by magnetic reconnection and episodic escapes of large plasmoids.
Over the course of the simulation ($ct/L_x \simeq 4.5$), the magnetic energy of the region $\mathcal{R}$ decreases by $\simeq 40\%$, while the particle energy decreases only by about $\simeq 15\%$.
At the same time, we measure a large net influx of electromagnetic energy (accumulating to $\simeq 1.5\,\mathcal{E}_{\rm B,0}$), mainly through the top/bottom boundaries of the region $\mathcal{R}$, and even larger net outflow of particle energy, mainly through the left/right boundaries.
The net energy outflow (particle minus electromagnetic) through the region boundaries amounts to $\simeq 0.25\,\mathcal{E}_{B,0}$ of the initial magnetic energy, which is slightly less than the particle energy lost to the synchrotron radiation ($\simeq 0.3\,\mathcal{E}_{B,0}$).
Accounting for all these energy components and flows, the total energy in $\mathcal{R}$ is conserved at the $\sim 0.1\%$ level.

%\clearpage
\subsection{Energy distributions of particles and photons}
\label{sec_res_spectra}

\begin{figure}
\includegraphics[width=\columnwidth]{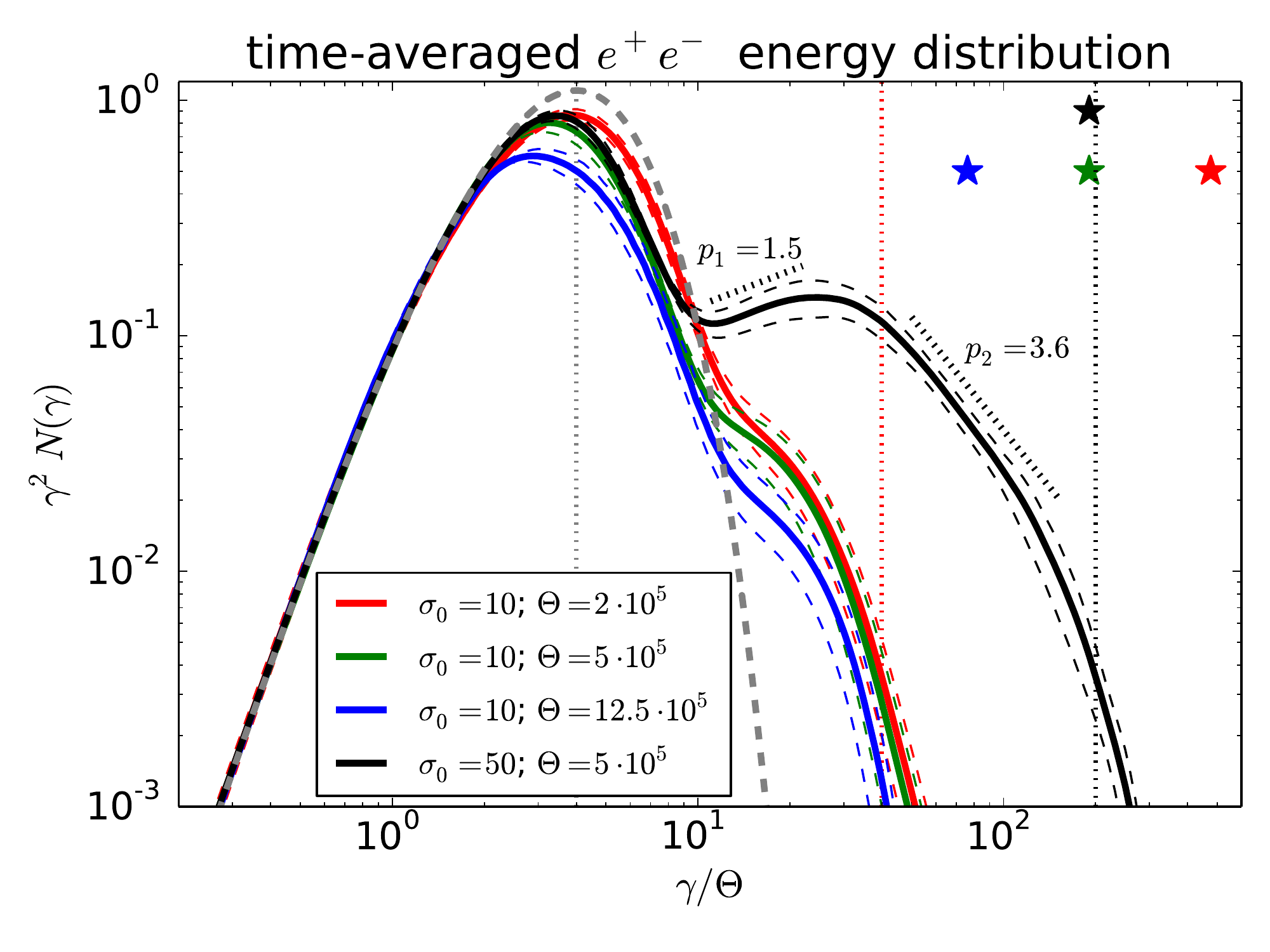}
\caption{Energy distributions of all particles contained in the simulation domain, compared for the 4 main simulations (higher $\Theta$ corresponds to more efficient cooling).
For each simulation, the distribution is averaged over simulation time, excluding the initial stage ($ct/L \lesssim 0.85$), in the space of flux logarithm.
The thin dashed lines indicate the corresponding standard deviation values.
The thick grey dashed line represents the initial Maxwell-J{\"u}ttner distribution for $\Theta \gg 1$.
The distributions are presented in arbitrary units, they are normalised to match the low-energy sections.
The vertical dotted lines indicate the characteristic values of $\gamma/\Theta$: $4$ (the $\gamma^2N(\gamma)$ peak for the initial background particles; grey), $4\times 10$ (red) and $4\times 50$ (black).
The oblique black dotted lines indicate two power-law slopes $p$ ($N(\gamma)\propto \gamma^{-p}$) along the $\sigma_0 = 50$ distribution.
The four stars indicate the values of $\gamma_{\rm rad}/\Theta$ for each simulation.}
\label{fig:spec_ele}
\end{figure}

\begin{figure}
\includegraphics[width=\columnwidth]{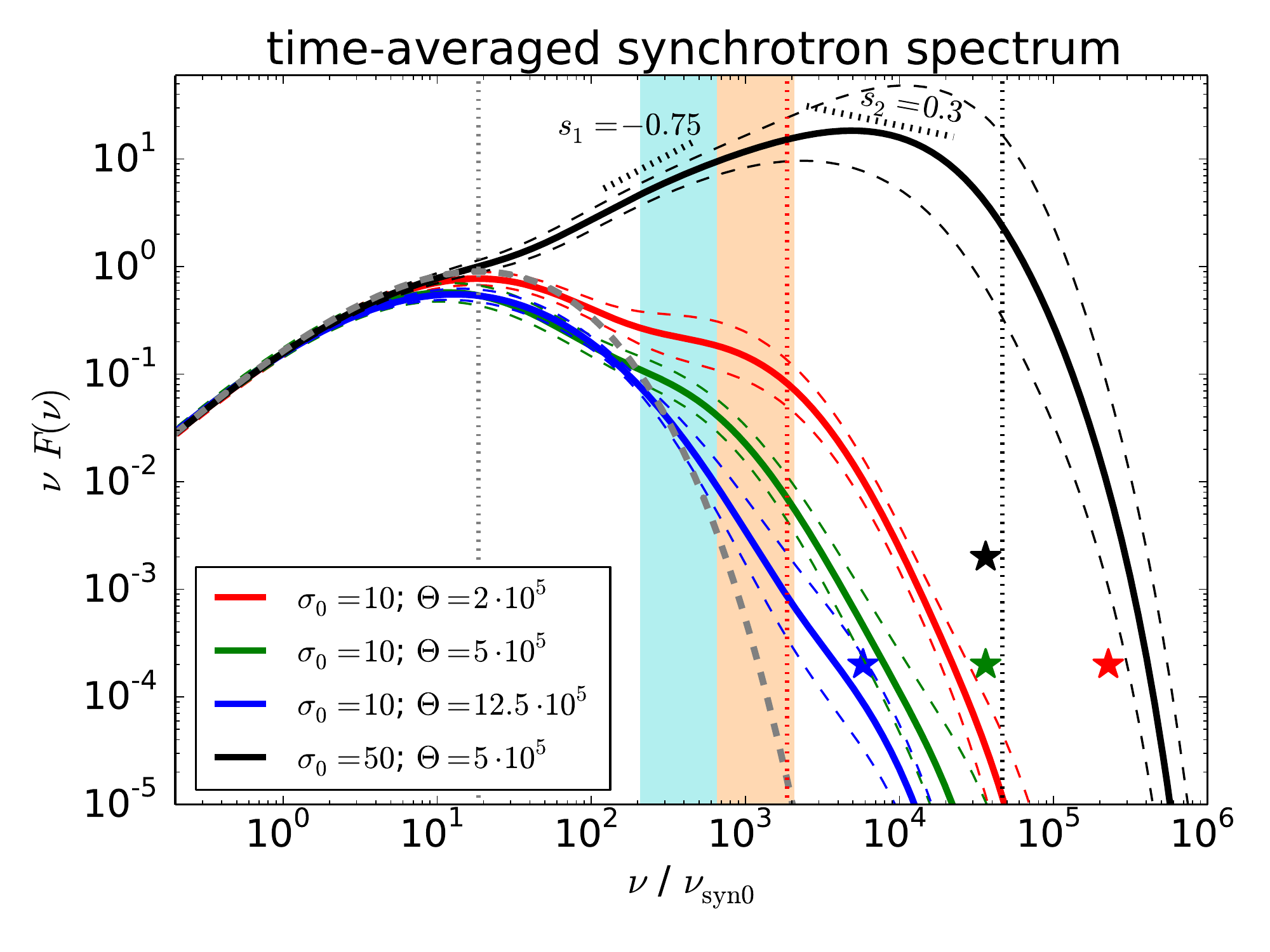}
\caption{Isotropic spectra of the synchrotron radiation emitted across the simulation domain, compared for the 4 main simulations. For each simulation, the distribution is averaged over simulation time, excluding the initial stage ($ct/L \lesssim 0.85$), in the space of flux logarithm. The thin dashed lines indicate the corresponding standard deviation values.
The thick grey dashed line represents the synchrotron spectrum of the initial Maxwell-J{\"u}ttner distribution.
The frequencies are normalised to the characteristic synchrotron frequency $\nu_{\rm syn0}$ defined in Eq. (\ref{eq_nu_syn0}). The distributions are presented in arbitrary units, they are normalised to match the low-frequency sections.
The vertical dotted lines indicate the characteristic values of $\nu/\nu_{\rm syn0}$: $19$ (the $\nu F(\nu)$ peak for the initial background particles; grey), $19\times 10^2$ (red) and $19\times 50^2$ (black).
The oblique black dotted lines indicate two power-law slopes $s = (3-p)/2$ ($\nu F_\nu \propto \nu^{-s}$) that would be expected for the corresponding power-laws $p_1,p_2$ marked in Figure \ref{fig:spec_ele}.
The four stars indicate the values of MHD synchrotron frequency limit $\nu_{\rm syn,max}/\nu_{\rm syn0} = (\gamma_{\rm rad}/\Theta)^2$ for each simulation.
The cyan and orange stripes indicate the frequency bands from which the lightcurves shown in Figure \ref{fig:lightcurves_xtmap_Esyn} were extracted.}
\label{fig:spec_syn}
\end{figure}

\begin{figure}
\includegraphics[width=\columnwidth]{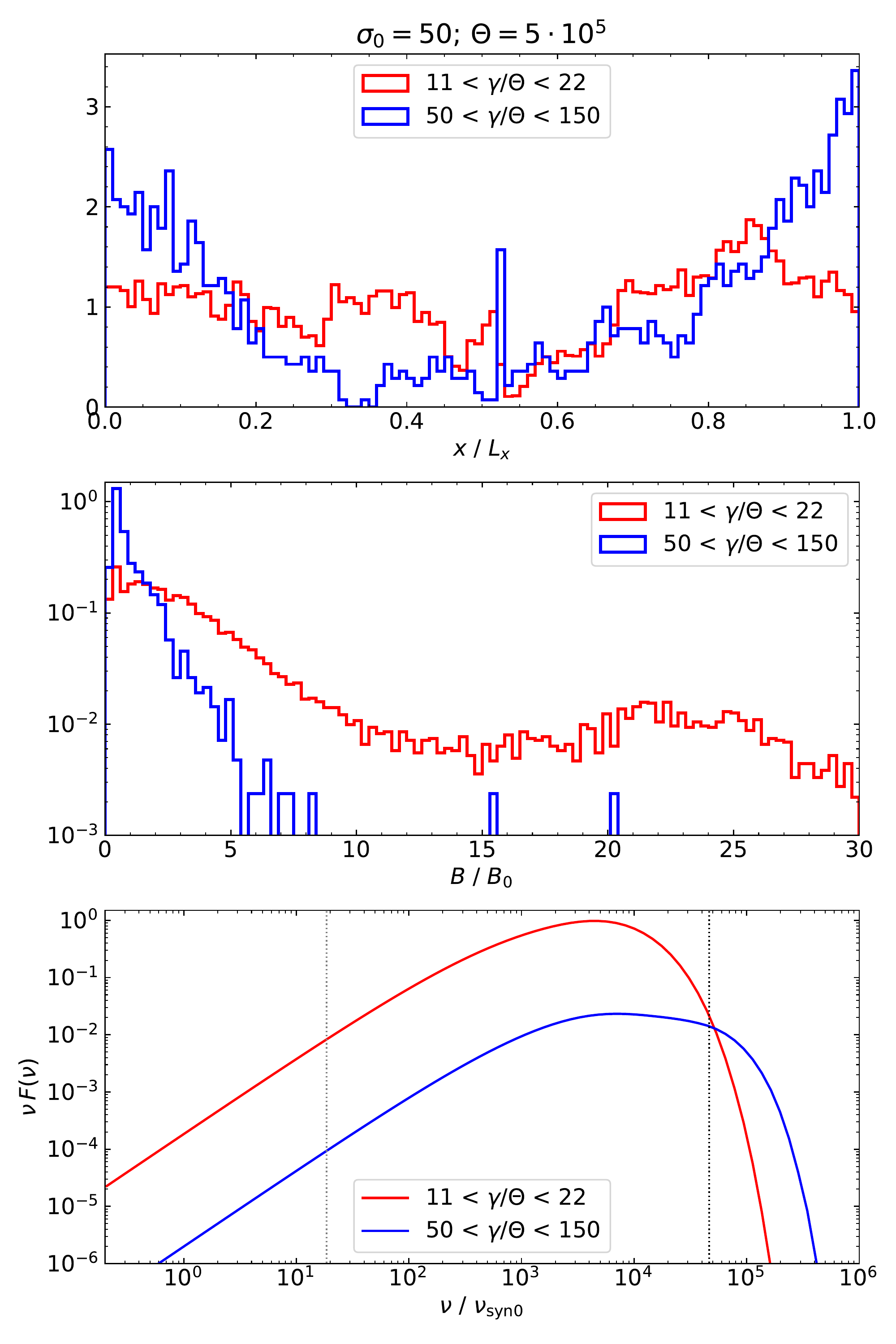}
\caption{Analysis of individual tracked particles for the simulation {\tt s50Tm}.
Particles are selected over two energy ranges --- $11 < \gamma/\Theta < 22$ (red) and $50 < \gamma/\Theta < 150$ (blue) --- corresponding to the two power-law sections indicated in Figure \ref{fig:spec_ele}.
The top panel compares their normalised distributions along coordinate $x$;
the middle panel compares their normalised distributions over magnetic field strength $B$;
the bottom panel compares their contributions to the isotropic synchrotron SED (arbitrary units).
The distributions are averaged over multiple simulation timesteps for $ct/L_x > 0.85$.
The vertical dotted lines in the bottom panel correspond to those in Figure \ref{fig:spec_syn}.}
\label{fig:orbits_distr}
\end{figure}

Figure \ref{fig:spec_ele} shows the energy distributions of all particles: electrons and positrons.
For each simulation, it is averaged over a period of time that excludes only the initial stage ($ct/L \lesssim 0.85$). In all studied cases, the particle energy distributions established after the initial period show no significant evolution in time. As energetic particles escape across the open left/right boundaries, other particles are energised across the current layer, and are subject to radiative energy losses within the plasmoids. The balance between these processes is maintained regardless of the efficiency of radiative cooling.
In all studied cases, a small fraction of particles reach energies of $\gamma_{\rm cutoff} = 4\sigma_0\Theta$ established as a cutoff energy in a previous study of non-radiative Harris-layer reconnection within periodic boundaries \citep{2016ApJ...816L...8W}.
In the case of $\sigma_0 = 10$, we find only a minor effect of radiative cooling in limiting the high-energy excess for $\Theta \gtrsim 10^6$.
In the case of $\sigma_0 = 50$, the high-energy component can be described as a broken power-law with a hard slope of $p_1 \simeq 1.5$ extending up to $\gamma \simeq 25\Theta$ and a soft tail of $p_2 \simeq 3.6$ extending up to $\gamma \simeq 150\Theta$.
In that case we also have $\gamma_{\rm cutoff} \simeq \gamma_{\rm rad}$.

Figure \ref{fig:spec_syn} shows the spectral energy distributions (SED) $\nu F_\nu$ of the synchrotron emission produced by all particles in all directions, averaged over the same periods of time as the particle energy distributions presented in Figure \ref{fig:spec_ele}.
In the case of $\sigma_0 = 10$, the SED are dominated by the contribution from low-energy particles peaking around $\nu \simeq 19\nu_{\rm syn0}$, with a high-frequency excess extending beyond a characteristic value of $\nu_{\rm cutoff} \simeq 19\sigma_0^2\nu_{\rm syn0}$.
The level of this high-frequency excess increases with decreasing gas temperature $\Theta$,
which means that radiative cooling suppresses the high-frequency radiation component more clearly than it affects the high-energy particle tail.
In the case of $\sigma_0 = 50$, the SED is strongly dominated by the contribution from energetic particles with the maximum photon energies consistent with a cutoff at $\nu_{\rm cutoff} \simeq 19\sigma_0^2\nu_{\rm syn0}$, which coincides with the radiation reaction limit $\nu_{\rm syn,max} = (\gamma_{\rm rad}/\Theta)^2\nu_{\rm syn0}$.
We note that the SED shape around its peak is not described by a broken power-law corresponding directly to that indicated in the electron distribution (with slopes $\nu F_\nu \propto \nu^{-s}$; $s = (p-3)/2$; and characteristic frequencies $\nu_i/\nu_{\rm syn0} \simeq (\gamma_i/\Theta)^2$).
This is because the extent of the electron energy distribution that can be described as a broken power-law is too short to result in a broken power-law photon spectrum when folded with the synchrotron kernel.

In order to clarify the connection between the electron energy distribution and synchrotron SED in the case of $\sigma_0 = 50$, we analysed a sample of individually tracked particles.
We selected particles over two energy ranges: (1) a medium-energy range $11 < \gamma/\Theta < 22$, corresponding to the hard power-law section of index $p_1 \simeq 1.5$; and (2) a high-energy range $50 < \gamma/\Theta < 150$, corresponding to the soft power-law section of index $p_2 \simeq 3.6$.
In Figure \ref{fig:orbits_distr} we show the distributions of these particles along coordinate $x$ and over local magnetic field strength $B$, as well as their contributions to the synchrotron SED, taking into account accurate local electromagnetic fields felt by each particle.
These distributions are averaged over multiple simulation timesteps for $ct/L_x > 0.85$.
For the medium-energy particles, we find that their distribution along $x$ is fairly uniform, and their distribution over $B$ is broad, with some particles found in strongly amplified magnetic field $B > 10B_0$ characteristic for the plasmoid cores.
For the high-energy particles, we find that they are clearly concentrated towards the left/right boundaries, and that they are found almost exclusively in magnetic fields of moderate strength $B < 5B_0$.
In addition, we observe that for individual simulation timesteps, the medium-energy particles are clearly concentrated within the plasmoids, while the high-energy particles are diffused over $x$.
The medium-energy particles dominate the synchrotron SED for most frequencies, except the highest values of $\nu > 5\times 10^4\nu_{\rm syn0}$.
At low frequencies, the relative contribution of the medium-energy particles is $\simeq 60$ times higher than that of the high-energy ones.

%\clearpage
\section{Discussion}
\label{sec_disc}

Our results are consistent with the basic picture of steady-state relativistic plasmoid reconnection that has been established since the work of \cite{2016MNRAS.462...48S}. The open boundaries allow to evacuate the reconnected plasma and develop unimpeded reconnection outflows that reach relativistic bulk velocities of the order of the background Alfven velocity. In the centrally positioned active magnetic X-point, plasmoids are generated spontaneously over a wide range of sizes.
The smallest discernible plasmoids, with the width of order $\sim 20\rho_0$ (deep blue lines in Figures \ref{fig:plasmoids_widths}, \ref{fig:plasmoids_temperatures} and \ref{fig:plasmoids_Esyn}) are rapidly accelerated to relativistic speeds, their lifetimes are about $\sim 0.2L_x/c$.
The largest plasmoids, with the width of order $\sim (150-200)\rho_0$ (brown lines in Figures \ref{fig:plasmoids_widths}, \ref{fig:plasmoids_temperatures} and \ref{fig:plasmoids_Esyn}) accelerate slowly, reaching mildly relativistic speeds $\sim c/2$ only after $\sim 1.5L_x/c$.
This is a confirmation of the anticorrelation between growth and acceleration of plasmoids identified by \cite{2016MNRAS.462...48S}.

An important consequence of the relation between plasmoid sizes and their acceleration length scale is that it is not possible to contain the bulk acceleration of large plasmoids within a domain with open boundaries. We have found that as we increase the size of the simulation domain, given proportionally more simulation time, we obtain ever larger plasmoids that require ever more acceleration time.
In other words, we are unable to isolate the bulk acceleration of plasmoids from the domain boundaries and obtain a coasting phase of uniformly Alfvenic reconnection outflows.

A similar problem applies to the production of synchrotron light curves.
In our simulations, even for the highest plasma temperatures investigated, we were not able to contain synchrotron emission from the largest plasmoids within the domain boundaries.
As we show in Figures \ref{fig:plasmoids_temperatures} and \ref{fig:plasmoids_Esyn}, even though in the case of short cooling length $l_{\rm cool} \ll L_x$ the plasmoid cores indeed undergo efficient radiative cooling within the $L_x/c$ time scale, the total synchrotron emission of plasmoid cores and layers is not decreasing significantly.

The nominal cooling time scale cannot be reduced indefinitely by a further increase of the particle distribution temperature. We found that, with unrestricted radiation reaction, already for $\Theta = 1.25\times 10^6$, the particle densities in the cores of the largest plasmoids increase to the level at which the Debye length eventually becomes unresolved $\lambda_{\rm D} = (\Theta m_{\rm e}c^2/4\pi ne^2)^{1/2} < {\rm d}x$, which leads to the development of numerical electrostatic instabilities.
Our ad hoc restriction of radiation reaction (see a footnote preceding Eq. \ref{eq:Psynchrotron}) helps to avoid developing these instabilities. We note that a similar restriction of radiation physics (pair creation) in the plasmoid cores has been applied by \cite{2019ApJ...877...53H}.

\cite{2019MNRAS.482...65C} calculated lightcurves of synchrotron and inverse Compton emission by post-processing the results of non-radiative PIC simulations of relativistic steady-state reconnection of \cite{2016MNRAS.462...48S}. Their radiation transfer model is based on several assumptions that can be verified by the results of our radiative PIC simulations. One of their key assumptions is that the intrinsic structure of plasmoids is not important and can be approximated by using their average parameters that in addition are constant in time. Our results suggest that the synchrotron emissivity is strongly concentrated in the central parts of the plasmoids (see the bottom panel of Figure \ref{fig:xymaps}), and that in the radiatively efficient regime the plasmoid cores undergo significant time evolution with systematic increase of plasmoid core density and peak magnetic field strength (Figure \ref{fig:plasmoids_widths}). We suggest that small plasmoids and the cores of large plasmoids are important for understanding the production of rapid radiation flares. Investigation of these structures is also the most challenging from the numerical perspective.

Our study suggests that properly resolving the cores of large plasmoids will be critical for understanding the radiative signatures of plasmoid reconnection. Recent non-radiative PIC simulations of relativistic reconnection demonstrated an important role of large plasmoids in extending the high-energy tail of the particle energy distribution along a power-law of slope $\simeq 2$ \citep{2018MNRAS.481.5687P}.
However, taking into account radiative cooling, which is expected to be particularly strong in the plasmoid cores, the maximum energy achievable in the plasmoids may be significantly limited.

Plasmoid mergers have been suggested previously to be important for particle acceleration in relativistic reconnection, based on the results of relatively modest PIC simulations of reconnection in periodic boundaries \citep{2015ApJ...815..101N}. Subsequent numerical studies emphasised a more fundamental role of magnetic X-points as a crucial first step for particles that eventually achieve the highest energies \citep{2019ApJ...879L..23G,2019ApJ...884...57B}.
Here we would like to point out that even if plasmoid mergers may not dominate particle acceleration in non-radiative reconnection, they are important for the production of transient radiation signals.
Even if major head-on collisions of large plasmoids are rare events, tail-on collisions of unequal plasmoids, arguably more frequent events due to the growth-acceleration anti-correlation, can be responsible for most of the sharpest features observed in the resulting lightcurves.
Recent results of \cite{2020MNRAS.492..549C} show that large plasmoids can attract many small plasmoids originating on both sides of their trajectory, enhancing the rates of both tail-on and head-on mergers.

Our results also demonstrate the co-existence of plasmoids and minijets in the same reconnection layer. We find minijets persisting in the gaps forming between plasmoids, plasmoids are able to slide along a minijet without causing much disturbance, and a minijet reforms behind a passing plasmoid. The structure of minijets found in our simulations is qualitatively very similar to the analytical model of \cite{2005MNRAS.358..113L}.
Although the minijets contain some highly energetic particles, their contribution to the observed radiative signatures appears to be very weak.
There may be two reasons behind this: (1) the minijets are characterised by much lower particle density than the plasmoids, (2) energetic particles propagating along the minijets show only weak radiative energy losses due to weak perpendicular magnetic field component.\footnote{Radiative cooling in the minijets might be stronger if the guide field component $B_z$ were included.}

In the high-magnetisation case of $\sigma_0 = 50$, we found that the particle energy distribution is maintained in the form of a broken power-law with a hard power-law slope $p_1 \simeq 1.5$ breaking around $\gamma \simeq 25\Theta$ into a soft power-law slope $p_2 \simeq 3.6$.
The hard power-law slope is consistent with the results of PIC simulations of non-radiative highly relativistic Harris-layer reconnection in periodic boundaries \citep{2014ApJ...783L..21S,Guo14,2016ApJ...816L...8W}.
The soft power-law slope is reminiscent of that seen in the recent PIC simulations of relativistic reconnection with strong inverse-Compton cooling, both in periodic boundaries \citep{2019MNRAS.482L..60W} and in the open ones \citep{2019arXiv190808138S}.
In the latter work, the electron energy distribution has been decomposed into contributions from particles accelerated at different sites, which suggests that the soft power-law arises from a combined action of primary X-points of the main reconnection layer, secondary X-points formed by merging plasmoids, and unstructured outflows (the minijets).
This is consistent with our analysis of the distribution of individual tracked particles belonging to the energy range of $50 < \gamma/\Theta < 150$.
We found that these particles are strongly concentrated towards the boundaries, and are found exclusively outside the plasmoid cores at normal magnetic field strengths $B < 5 B_0$.
Typical examples of such particles are Particles \#3 and \#4 presented in Figures \ref{fig:Particle_xt_track} and \ref{fig:Particle_track}.
Low relative numbers and diffuse spatial distribution implies that these particles are not important to the overall synchrotron SED (except for the very highest frequencies) nor to the production of rapid radiation flares.

\section{Conclusions}
\label{sec_conc}

We presented the results of the first kinetic simulations of relativistic magnetic reconnection (RMR) within open boundaries that enable steady-state plasmoid reconnection and including the synchrotron radiation reaction.
We confirm the general picture of steady-state relativistic plasmoid reconnection established by \cite{2016MNRAS.462...48S} and subsequent works.
We find that synchrotron emission of plasmoids cannot be contained within open boundaries.
The cores of large plasmoids are the main sites of synchrotron emission, their particle densities are significantly enhanced due to radiative pressure losses.
Rapid flares of synchrotron radiation can be produced by tail-on mergers between small/fast plasmoids with large/slow targets.
The plasmoids are also found to co-exist with the minijets that do not produce a lot of radiation due to their low particle densities. 
In the high-magnetisation case ($\sigma_0 = 50$), the energy distribution of accelerated particles can be described as a broken power-law with a hard medium-energy section produced mainly by particles accelerated in the plasmoids and a soft high-energy tail produced by diffuse particles accelerated in the minijets.

\section*{Acknowledgements}
We thank the Reviewer for helpful comments.
We acknowledge discussions with Beno{\^i}t Cerutti, Dimitrios Giannios and Maria Petropoulou.
The original version of the {\tt Zeltron} code was created by Beno{\^i}t Cerutti and co-developed by Gregory Werner at the University of Colorado Boulder (\url{http://benoit.cerutti.free.fr/Zeltron/}).
These results are based on numerical simulations performed at the supercomputer \emph{Prometheus} located at Cyfronet AGH, Poland (PLGrid grant {\tt recjose18}; PI: K. Nalewajko).
This work was supported by the Polish National Science Centre grant 2015/18/E/ST9/00580.

%----------------------------------------------------------------
%	REFERENCE LIST
%----------------------------------------------------------------

\end{document}